\documentclass[reprint,prb]{revtex4-1}
\usepackage{bm,epsfig,graphicx,color,amssymb,amsmath}

\begin{document}

\title{Electromigration of bivalent functional groups on graphene}

\author{Kirill A. Velizhanin}

\email{kirill@lanl.gov}

\address{Theoretical Division, Los Alamos National Laboratory, Los Alamos,
NM 87545, USA}

\author{Naveen Dandu}

\address{Department of Chemistry and Biochemistry, North Dakota State University,
Fargo, ND 58102, USA}

\author{Dmitry Solenov}

\email{d.solenov@gmail.com}
\altaffiliation{Present address: Naval Research Laboratory, 4555 Overlook Ave., SW Washington, District of Columbia 20375, USA}
\affiliation{National Research Council, National Academies, Washington, District of Columbia 20001, USA}

\begin{abstract}
Chemical functionalization of graphene holds promise for various applications ranging from nanoelectronics to catalysis, drug delivery, and nano-assembly. In many of these applications it is critical to assess the rates of electromigration - directed motion of adsorbates along the surface of current-carrying graphene due to the electron wind force. In this paper, we develop an accurate analytical theory of electromigration of bivalent functional groups (epoxide, amine) on graphene. Specifically, we carefully analyze various factors contributing to the electron wind force, such as lattice effects and strong scattering beyond Born approximation, and derive a simple analytical expression for this force. Further, we perform accurate electronic structure theory calculations to parameterize the obtained analytical expression. The obtained results can be generalized to different functional groups and adsorbates, e.g., alkali atoms on graphene.
\end{abstract}

\maketitle

\section{Introduction}

Graphene -- a single atom thick honeycomb lattice of carbon - is a unique material combining the superb mechanical, electronic, optical and thermal properties.\cite{Geim2007-183,Lee2008-385,Balandin2008-902} Nevertheless, ever since the first experimental isolation of graphene in 2004,\cite{Novoselov2004-666} there has been a growing interest in ``enhancement" of graphene by means of chemical functionalization.\cite{Boukhvalov2009-344205,Georgakilas2012-6156,Kuila2012-1061}  This is especially promising since unlike three-dimensional (3D) materials where chemical functionalization typically affects only surface properties, graphene is affected and modified in its entirety because of its truly 2D  ``all-surface'' nature. For instance, chemically functionalized graphene (e.g., graphene
oxide, reduced graphene oxide, graphane) can have electronic properties very different from those of pristine graphene, which holds promise in nanoelectronics and optics,\cite{Boukhvalov2008-4373,Loh2010-1015,Englert2011-279}
nonvolatile memory,\cite{Cui2011-6826} graphene-based nanoassemblies
for catalysis, photovoltaics and fuel cells applications.\cite{Kamat2010-520,Kamat2011-242}

The reliability of a functionalized graphene layer within a device is directly linked to the stability of its chemical functionalization against various external stimuli such as temperature, light, as well as thermal and electric currents. For example, functional groups can desorb from the graphene surface or participate in various on-surface chemical transformations causing the deterioration of the device characteristics with time.\cite{Zhou2013-2484,Dreyer2010-228,Kim2012-544,Pembroke2013-138} Furthermore, even if functional groups remain chemically intact and do not desorb, characteristics of a graphene-based device can still change with time provided functional groups can {\em move} along the surface.  For example, this motion can break the ordering of functional groups, thus strongly affecting the electronic structure of graphene (e.g., bandgap).\cite{Robinson2010-3001,Abanin2010-086802}

Motion of functional groups along the surface of graphene occurs either via diffusion\cite{Suarez2011-146802} or via drift when, e.g., an electric current is present.\cite{Solenov2012-095504} The latter phenomenon - electromigration - originates from scattering of in-graphene charge carriers by a functional group. The force resulting from the scattering-mediated momentum transfer is universally referred to as the {\em electron wind force}.\cite{Sorbello1997-159} We have recently demonstrated theoretically a surprising efficiency of electromigration of bivalent functional groups on graphene.\cite{Solenov2012-095504} Therefore, on one hand, thorough understanding of this phenomenon is critical in order to assess the rate of aging of functionalized graphene within current-carrying devices. On the other hand, the electromigration can be exploited as an efficient and easily controllable means to drive the mass transport along the surface of graphene. This can be of use in nano-assembly\cite{Yang2008-124709} and drug delivery applications.\cite{Voloshina2011-220}

In this paper, we develop an accurate model for electromigration of bivalent adsorbates on graphene, focusing on epoxy ($-{\rm O}-$) and amino ($-{\rm NH}-$) groups. We significantly expand on results of our previous work\cite{Solenov2012-095504} by (i) systematically analyzing  various equilibrium and non-equilibrium contributions to the interaction between these functional groups and graphene, (ii) evaluating the driving force of electromigration ``to all orders", (iii) investigating the graphene crystal lattice effects on electromigration, and (iv) performing accurate electronic structure theory calculations to carefully parameterize the developed model.

\begin{figure}
\includegraphics[width=3.3in]{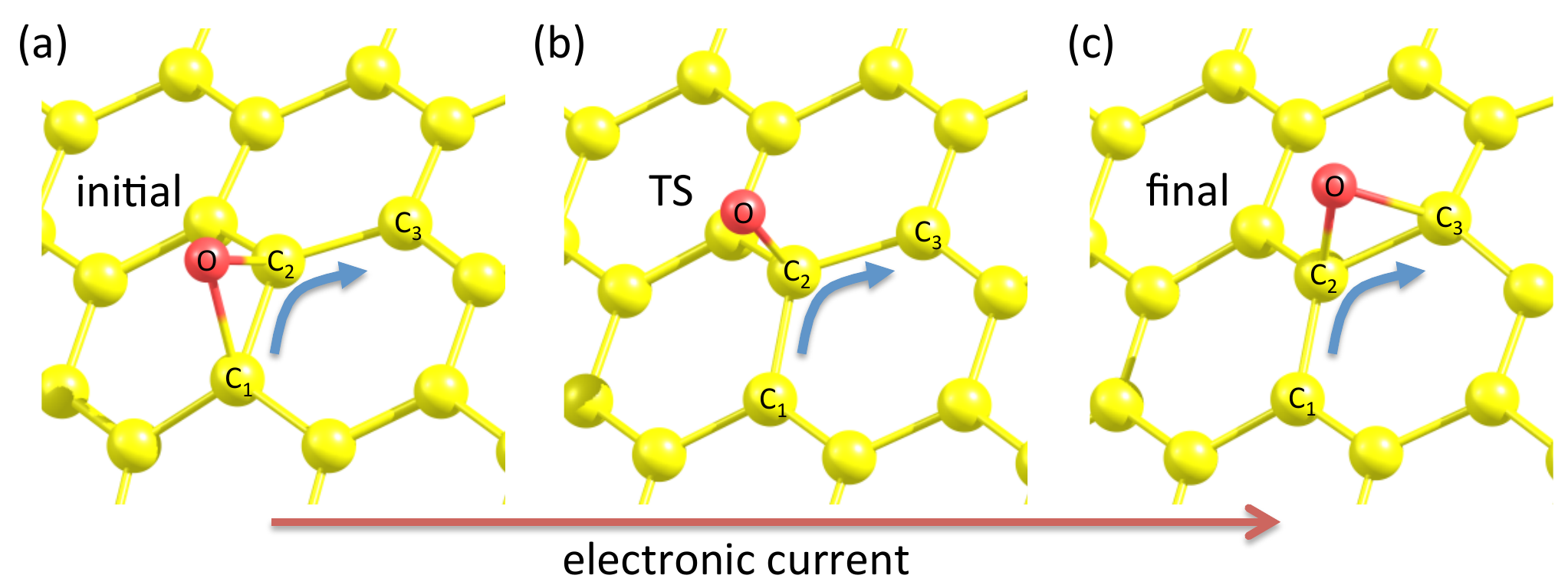}
\caption{\label{fig:Schematic} Electromigration of a bivalent functional group (epoxy) on graphene. An oxygen atom jumps from the initial stable configuration (a) to the final stable configuration (c) via a hopping transition state (TS), (b). The blue arrow represents the elementally hopping trajectory. The functional group forms two covalent bonds with carbon atoms in the stable configurations: ${\rm OC_1}$ and ${\rm OC_2}$ in the initial state and ${\rm OC_2}$ and ${\rm OC_3}$ in the final one. Only a single covalent bond, ${\rm OC_2}$, is formed in the transition state. Red horizontal arrow shows the direction of electronic current that defines the preferential direction of hopping.
}
\end{figure}

Figure~\ref{fig:Schematic} depicts an elementary hopping trajectory of a single oxygen atom (epoxy group) on graphene. 
The activation energy of this hopping, i.e., the energy difference between the transition state (TS) and initial (or final) states has been recently shown to drop from $\sim$0.7-0.8~eV for neutral graphene to $\sim$0.15~eV for $n$-doped graphene. Diffusion becomes very fast in the latter case with a diffusion coefficient as high as $\sim$10$^{-6}$~cm$^{2}$/s .\cite{Suarez2011-146802} A secondary amine ($-{\rm NH}-$) binds to graphene very similarly, i.e., it forms two covalent bonds with graphene in the stable configuration and only a single one in the TS. 

The paper is organized as follows. Sec.~\ref{sec:General} is devoted to description of the general theory of electron wind force. Specifically, it deals with the careful separation of equilibrium and non-equilibrium, as well as conservative and non-conservative contributions to interaction between graphene and functional groups. An expression for the magnitude of electron wind force acting on bivalent functional groups on graphene is derived in Sec.~\ref{sec:graphene}. Sec.~\ref{sec:res} presents the model parameterization by means of electronic structure theory calculations and the numerical results for electromigration efficiency of epoxy and amino groups on graphene. Concluding remarks are given in Sec.~\ref{sec:concl}. 

\section{General theory of electron wind force}\label{sec:General}

This section is devoted to the detailed description of the electron wind force that pushes functional groups along a substrate (graphene) in a non-equilibrium situation. Specifically, in this work we go beyond a simple jellium model derivation of Ref.~\onlinecite{Solenov2012-095504} and take the substrate lattice into account. Due to the presence of this lattice, the force acting on a functional group can be non-zero at different positions with respect to the lattice sites even when no current is flowing through the system. Because of this, a separation of conservative and non-conservative (wind force) components of the total force becomes non-trivial. It will be shown shortly that this separation becomes unambiguous when the force is {\it averaged} over trajectories connecting two equivalent crystallographic points. This allows us to introduce an effective local electron wind force.

The derivations to follow are sufficiently general and applicable to not just specific bivalent functional groups that are of interest in this work, but also to generic {\em adsorbates} on graphene. Thus the term ``adsorbate" and ``functional group" will be used interchangeably below.

\subsection{Microscopic force acting on an adsorbate}

Interaction of an adsorbate with a substrate (e.g., graphene) results in a microscopic force acting on the adsorbate. We denote an effective electrostatic potential of the adsorbate by
\begin{equation}
U_{{\bf R}}({\bf x})=U({\bf x}-{\bf R}),\label{eq:Apot}
\end{equation}
so that the Hamiltonian term encoding the interaction of the adsorbate
with the substrate is given by $\hat{H}_{int}=\int d{\bf x}\, U_{{\bf R}}({\bf x})\hat{\rho}({\bf x})$,
where $\hat{\rho}({\bf x})$ is the operator of spatially-resolved
charge density of the substrate. The position of the adsorbate is
denoted by ${\bf R}$. We focus on a non-magnetic case and suppress spin degrees of freedom unless stated otherwise. Further, we assume the adsorbate
to be \emph{structureless}, i.e., it is represented by
a local single-particle potential, $U_{{\bf R}}({\bf x})$, with no
frequency dependence. We emphasize here that $\hat{\rho}({\bf x})$
is the total charge density of the substrate in a sense that it incorporates
both electrons and nuclei of the substrate. Once the adsorbate-substrate
interaction is defined in the operator form, the force acting on the
adsorbate from the substrate in the adiabatic regime can be calculated
as \cite{Sorbello1997-159}
\begin{equation}
{\bf F}=-\int d{\bf x}\:\left[\nabla_{{\bf R}}U_{{\bf R}}({\bf x})\right]\rho({\bf x}),\label{eq:total_force}
\end{equation}
where $\rho(x)$ is the quantum-mechanical average of $\hat{\rho}(x)$
over the exact state of the adsorbate-substrate system. This averaging over the exact state of the entire system naturally accounts for the effects of the adsorbate-substrate interaction
``to all orders'', and not just perturbatively. Furthermore, the state of the system can be a
non-equilibrium one, e.g., with DC electric current flowing through
the substrate. 

In the thermodynamic equilibrium, Eq.~(\ref{eq:total_force}) is
just a representation of the Hellmann-Feynman theorem, which relates
the force to the variation of the total energy of the system.\cite{Hellmann1937,Feynman1939-340}
The derivation is more involved in the non-equilibrium case. However, the result of this derivation remains
formally the same as in the equilibrium case, i.e., Eq.~(\ref{eq:total_force})
still holds true.\cite{Sorbello1997-159}

In this work, a non-equilibrium electronic state of the substrate
before the attachment of the adsorbate is specified through the populations
of single-electron states
\begin{equation}
f_{i}=f_{i}^{eq}+\delta f_{i},\label{eq:neq_popul}
\end{equation}
where $f_{i}^{eq}$ is the equilibrium Fermi-Dirac population of state
$i$, and $\delta f_{i}$ is the deviation of the population of this
state from the equilibrium one. We expand the charge density of the substrate
in the presence of the adsorbate, $\rho({\bf x})$,
into a double series with respect to $U_{{\bf R}}({\bf x})$
and $\delta f_{i}$,
\begin{equation}
\rho({\bf x})=\sum_{m=0}^{\infty}\sum_{n=0}^{1}\rho_{m,n}({\bf x}),\label{eq:rho_exp}
\end{equation}
where one has (symbolically) $\rho_{m,n}\propto\left(U_{{\bf R}}\right)^{m}\left(\delta f\right)^{n}$.
Substituting Eq.~(\ref{eq:rho_exp}) into Eq.~(\ref{eq:total_force})
one obtains the expansion of the adsorbate-substrate force as ${\bf F}=\sum_{m,n}{\bf F}_{m,n}$,
where
\begin{equation}
{\bf F}_{m,n}=-\int d{\bf x}\,\left[\nabla_{{\bf R}}U_{{\bf R}}({\bf x})\right]\rho_{m,n}({\bf x}).\label{eq:F_exp}
\end{equation}

In this work, we treat the deviation from equilibrium within the
linear response approximation so that $\rho$ is at most linear with
$\delta f$. Accordingly, it is convenient to use more explicit subscripts:
``$eq$" for $n=0$ and ``$neq$" for $n=1$. We
first consider all the equilibrium expansion terms, i.e., those with
$n=eq$. These terms sum up into the equilibrium contribution to the
force acting on the adsorbate
\begin{equation}
{\bf F}_{eq}=-\int d{\bf x}\,\left[\nabla_{{\bf R}}U_{{\bf R}}({\bf x})\right]\rho_{eq}({\bf x}),\label{eq:force_eq}
\end{equation}
where $\rho_{eq}({\bf x})=\sum_{m=0}^{\infty}\rho_{m,eq}({\bf x})$. The corresponding equilibrium adsorbate-substrate
interaction potential can be introduced as
\begin{equation}
V_{eq}({\bf R})=\int d{\bf x}\, U_{{\bf R}}({\bf x})\rho_{eq}({\bf x}),\label{eq:eq_inter}
\end{equation}
so that ${\bf F}_{eq}=-\nabla_{\bf R} V_{eq}({\bf R})$ is {\em conservative}.
The local energy minima of the potential energy landscape, generated
by $V_{eq}({\bf R})$, determine the equilibrium configurations of
the adsorbate-substrate system. Saddle points of this landscape define
transition states of hopping diffusion of adsorbates along the substrate.
In Sec.~\ref{sec:res} the potential given by Eq.~(\ref{eq:eq_inter})
will be obtained for covalently-bound adsorbates on graphene via
electronic structure theory calculations.

Of all the non-equilibrium terms in Eq.~(\ref{eq:F_exp}), i.e.,
those linear in $\delta f$, we consider here only the first two, ${\bf F}_{0,neq}$ and ${\bf F}_{1,neq}$, and estimate the contribution
of higher-order terms later [see Eq.~(\ref{eq:UC_unitless}) and App.~\ref{app:strong}]. The former is
\begin{equation}
{\bf F}_{0,neq}=-\int d{\bf x}\,\left[\nabla_{{\bf R}}U_{{\bf R}}({\bf x})\right]\rho_{0,neq}({\bf x}).\label{eq:force_01}
\end{equation}
Since $\rho_{0,neq}({\bf x})$ does not depend on $U_{{\bf R}}$,
this expression is an exact differential so it is possible to introduce
$V_{0,neq}({\bf R})$ so that ${\bf F}_{0,neq}=-\nabla_{{\bf R}}V_{0,neq}({\bf R})$.
Therefore, this contribution is conservative and can be considered as a non-equilibrium
correction to the equilibrium adsorbate-substrate interaction
potential, Eq.~(\ref{eq:eq_inter}). 

The second non-equilibrium term considered, i.e., ${\bf F}_{1,neq}$,
can be understood using the following auxiliary considerations. The
linear response of the substrate charge density to the adsorbate potential
in a general non-equilibrium situation can be written as
\begin{equation}
\rho'({\bf x})=\int d{\bf x}\,\chi({\bf x},{\bf x}')U_{{\bf R}}({\bf x}'),\label{eq:rho_linear}
\end{equation}
where the non-equilibrium Kubo response function of the substrate
to the static external perturbation is denoted by $\chi({\bf x},{\bf x}')$. Eq.~(\ref{eq:rho_linear}) is equivalent to the first-order Born approximation in the scattering theory.\cite{Economou1990}

Expanding response function $\chi({\bf x},{\bf x}')$ into powers of $\delta f$ up to
linear terms, one obtains
\begin{equation}
\chi({\bf x},{\bf x}')=\chi_{eq}({\bf x},{\bf x}')+\delta\chi({\bf x},{\bf x}'),\label{eq:chi_expans}
\end{equation}
where the last r.h.s. term is linear with respect to $\delta f$.
The contribution of the first r.h.s. term of this expression to the
adsorbate-substrate force has been already accounted for within ${\bf F}_{eq}$
(as ${\bf F}_{1,eq}$). Then, $\delta\chi({\bf x},{\bf x}')$ determines
${\bf F}_{1,neq}$. We consider the case when
$\delta f$ is non-negligible only around the Fermi surface. At this condition, it is natural to assume that $\delta\chi({\bf x},{\bf x}')$ is mostly defined by the electrons near the Fermi surface of graphene, i.e., $\delta\chi({\bf x},{\bf x}')\approx\delta\chi_{e}({\bf x},{\bf x}')$,
where the non-equilibrium contribution to the electronic response
function of the substrate is denoted by $\delta\chi_{e}$.

The non-equilibrium electronic response function, $\delta\chi_{e}({\bf x},{\bf x}')$,
can in general be non-symmetric with respect to the permutation ${\bf x}\leftrightarrow{\bf x}'$ and, therefore, can be written as a sum of its symmetric and antisymmetric
parts as
\begin{equation}
\delta\chi_{e}({\bf x},{\bf x}')=\delta\chi_{e}^{+}({\bf x},{\bf x}')+\delta\chi_{e}^{-}({\bf x},{\bf x}'),
\end{equation}
where $\delta\chi_{e}^{\pm}({\bf x},{\bf x}')=\left[\delta\chi_{e}({\bf x},{\bf x}')\pm\delta\chi_{e}({\bf x}',{\bf x}')\right]/2$
and the $\pm$ superscript stands for the parity with respect to the coordinate
permutation. It is straightforward to see that the symmetric contribution
to ${\bf F}_{1,neq}$, given by
\begin{equation}
{\bf F}_{1,neq}^{+}=-\int d{\bf x}d{\bf x}'\,\left[\nabla_{{\bf R}}U_{{\bf R}}({\bf x})\right]\delta\chi_{e}^{+}({\bf x},{\bf x}')U_{{\bf R}}({\bf x}'),\label{eq:force_11_symm}
\end{equation}
forms an exact differential, and, therefore, is conservative. Similarly to Eq.~(\ref{eq:force_01}), ${\bf F}^+_{1,neq}$ is a non-equilibrium correction to the equilibrium force, Eq.~(\ref{eq:force_eq}).

The antisymmetric contribution
\begin{equation}
{\bf F}_{1,neq}^{-}=-\int d{\bf x}d{\bf x}'\,\left[\nabla_{{\bf R}}U_{{\bf R}}({\bf x})\right]\delta\chi_{e}^{-}({\bf x},{\bf x}')U_{{\bf R}}({\bf x}'),\label{eq:force_11_asymm}
\end{equation}
gives, in general, a non-conservative force since it does not form an exact differential.

Gathering all the contributions to the adsorbate-substrate force analyzed
so far, Eqs.~(\ref{eq:force_eq}), (\ref{eq:force_01}), (\ref{eq:force_11_symm})
and (\ref{eq:force_11_asymm}), and neglecting higher order non-equilibrium terms one obtains 
\begin{equation}
{\bf F}=\left\{ {\bf F}_{eq}+{\bf F}_{0,neq}+{\bf F}_{1,neq}^{+}\right\} _{c}+{\bf F}_{1,neq}^{-}.\label{eq:tforce}
\end{equation}
The curly brackets with the subscript $c$ emphasize the conservative
nature of the terms within. The last term in Eq.~(\ref{eq:tforce}) is a non-conservative
force driving the electromigration: the electron wind force.

\subsection{Electron wind force\label{sub:wind_force}}

Let us consider a situation where an adsorbate is hopping along a
certain trajectory between two crystallographically equivalent points
of the substrate lattice. These points are specified by the initial
and final positions of the adsorbate, ${\bf R}_{1}$ and ${\bf R}_{2}$,
respectively. The average force acting on the adsorbate along this
hopping trajectory is then given by
\begin{equation}
F_{w}=l_{12}^{-1}\int_{{\bf R}_{1}}^{{\bf R}_{2}}\left(d{\bf R}\cdot{\bf F}\right),\label{eq:Fw_org}
\end{equation}
where the integral is evaluated along the hopping trajectory and the
microscopic force, ${\bf F}$, is given by Eq.~(\ref{eq:tforce}).
The length of the trajectory is denoted by $l_{12}$. The sum of conservative
terms in Eq.~(\ref{eq:tforce}) can be considered as a gradient of
a certain scalar potential, $V({\bf R})$. This potential
has the same translational symmetry as the substrate lattice resulting in $V({\bf R}_{1})-V({\bf R}_{2})=0$. Then, the contribution of conservative terms into the averaged force is zero so Eq.~(\ref{eq:Fw_org})  reduces to
\begin{equation}
F_{w}=l_{12}^{-1}\int_{{\bf R}_{1}}^{{\bf R}_{2}}\left(d{\bf R}\cdot{\bf F}_{1,neq}^{-}\right).
\end{equation}
This force breaks the equivalence of ${\bf R}_{1}$
and ${\bf R}_{2}$ points within the substrate lattice, and, therefore,
can drive the directed hopping, i.e., migration, of adsorbates along the substrate. From what follows it will become clear why this
force can be called \emph{electron wind force} (hence, the subscript
$w$ - wind). For now, we just emphasize that the rigorous definition
of the electron wind force can only be done through averaging of the
position-dependent force over the hopping trajectory connecting two
crystallographically equivalent points. However, it is still possible (although ambiguous)
to introduce an effective \emph{local} wind force , ${\bf F}_{w}$, by choosing
any vector field that produces the correct averaging over hopping
trajectories.%
\footnote{The ambiguity in defining the electron wind force locally is reminiscent
of gauge freedom in electrodynamics. Indeed, one can add a gradient
of an arbitrary scalar function to the wind force defined locally.
If the chosen scalar function is periodic over the substrate unit
cell, its specific form does not affect the rigorous definition of
the wind force, i.e., the one done via averaging.%
} In our case an obvious choice is ${\bf F}_{1,neq}^{-}$
- the choice we will assume henceforth.

In what follows, we will neglect the conservative non-equilibrium corrections
to the equilibrium force, ${\bf F}_{eq}$. Furthermore, we
will show later that the higher order (with respect to $U_{{\bf R}}$)
corrections to the wind force are small compared to
the leading one, ${\bf F}_{1,neq}^{-}$, for the specific system considered
in this work (covalently bound adsorbates on graphene) . These two assumptions transform
Eq.~(\ref{eq:tforce}) into
\begin{equation}
{\bf F}=-\nabla_{{\bf R}}V_{eq}({\bf R})+{\bf F}_{w}.\label{eq:tforce_finale}
\end{equation}
Here, the first term determines the potential energy landscape where the diffusion of the
adsorbate occurs. The second term, ${\bf F}_w\equiv {\bf F}^-_{1,neq}$, breaks the translational
invariance of the lattice, thus pushing adsorbates along the substrate. 

The electronic response function is given by\cite{Giuliani2005}
\begin{equation}
\delta\chi_{e}({\bf x},{\bf x}')=\frac{e^2}{\hbar}\sum_{i,j}\frac{\delta f_{i}-\delta f_{j}}{\omega_{ij}+i\delta}\eta_{ij}({\bf x})\eta_{ji}({\bf x}'),\label{eq:chi_xx}
\end{equation}
where $e$ is the magnitude of the electron charge and $\eta_{ij}({\bf x})=\Psi_{i}^{*}({\bf x})\Psi_{j}({\bf x})$
is the transition density.  The normalized Bloch wave of electronic state $i$ is defined
as
\begin{equation}
\Psi_{i}({\bf x})=S^{-1/2}e^{i{\bf k}_{i}{\bf x}}u_{i}({\bf x}),\label{eq:wfunc}
\end{equation}
where $u_{i}({\bf x})$ is the Bloch (or cellular) function periodic
over a unit cell. The area of the substrate is denoted by $S$. Multi-index
$i$ specifies a certain single-electron state via all its quantum
numbers, such as wavenumber, ${\bf k}_{i}$, band index, $b_{i}$, etc. The energy of single-electron state $i$ is given by $\hbar\omega_i$, and therefore, the energy difference between states $i$ and $j$ is $\hbar\omega_{ij}=\hbar\omega_i-\hbar\omega_j$. The
explicit dependence of the response function $\delta\chi_{e}({\bf x},{\bf x}')$
on ${\bf x}$ and ${\bf x}'$ and not just on ${\bf x}-{\bf x}'$
comes from the local-field effects due to perturbation of the charge
density on the scale of a single unit cell.\cite{Schilfgaarde2011-081409}
Using the Sokhotski\textendash{}Plemelj identity, $\frac{1}{\omega+i\delta}=\mathcal{P}\frac{1}{\omega}-i\pi\delta(\omega)$
(where $\mathcal{P}$ stands for the Cauchy principal value), the
antisymmetric part of the response function can be written as 
\begin{align}
\delta\chi_{e}^{-}({\bf x},{\bf x}')&=\left[\delta\chi_{e}({\bf x},{\bf x}')-\delta\chi_{e}({\bf x}',{\bf x})\right]/2\nonumber\\
&=-i\frac{\pi e^2}{\hbar}\sum_{i,j}\left[\delta f_{i}-\delta f_{j}\right]\eta_{ij}({\bf x})\eta_{ji}({\bf x}')\delta(\omega_{ij}).\label{eq:chi_asymm}
\end{align}
Despite the imaginary unit in front, this expression is easily shown
to be real. Since a Bloch function, $u_{i}(x)$, is periodic over
a unit cell, its expansion into the discrete Fourier series yields
\begin{equation}
u_{i}^{*}({\bf x})u_{j}({\bf x})=\sum_{{\bf G}}B_{ij}({\bf G})e^{i{\bf G}{\bf x}},\label{eq:dFexp}
\end{equation}
where the Bloch function overlap coefficients are given by 
\begin{equation}
B_{ij}({\bf G})=s^{-1}\int_{s}d{\bf x}\, u_{i}^{*}({\bf x})u_{j}({\bf x})e^{-i{\bf G}{\bf x}}.\label{eq:B_exp}
\end{equation}
Here, the integral is evaluated over the area of a single unit cell,
$s$, and vector ${\bf G}$ runs through all possible translation
vectors of the reciprocal lattice of the substrate. Combining this
expression with Eqs.~(\ref{eq:chi_asymm}) and (\ref{eq:force_11_asymm})
and using the identity 
\begin{equation}
\int d{\bf x}\,\eta_{ij}({\bf x})U({\bf x})=S^{-1}\sum_{{\bf G}}B_{ij}({\bf G})U_{{\bf R}}({\bf k}_{i}-{\bf k}_{j}-{\bf G}),\label{eq:etaU_Fourier}
\end{equation}
one obtains 
\begin{align}
{\bf F}_{w}={\bf F}_{1,neq}^{-}&=\frac{\pi e^2}{\hbar S^{2}}\sum_{i,j}\delta(\omega_{ij})\left[\delta f_{i}-\delta f_{j}\right]\nonumber\\
&\times\sum_{{\bf G}}({\bf k}_{i}-{\bf k}_{j}-{\bf G})U_{{\bf R}}({\bf k}_{i}-{\bf k}_{j}-{\bf G})B_{ij}({\bf G})\nonumber\\
&\times\sum_{{\bf G}'}U_{{\bf R}}({\bf k}_{j}-{\bf k}_{i}-{\bf G}')B_{ji}({\bf G}').\label{eq:Fw_general}
\end{align}
This expression describes the process of electron scattering from
state $i$ to state $j$ (with the subsequent averaging over these
states), where the rate (probability per unit time) of each elementary
act of scattering is weighted with the momentum transfer factor, ${\bf k}_{i}-{\bf k}_{j}-{\bf G}$,
so that the net result is the force acting on the adsorbate due to
the scattering of itinerant electrons. Hence the name \emph{electron
wind force}. The two factors to the right of the first summation symbol,
i.e., $\delta(\omega_{ij})$ and $\delta f_{i}-\delta f_{j}$, guarantee
that the scattering is \emph{elastic} (i.e., it conserves energy)
and occurs only in the close proximity of the Fermi surface, respectively.
Below we briefly discuss a few scenarios of how Eq.~(\ref{eq:Fw_general})
can be used in various practical situations.

In the case of a free electron gas (i.e., jellium model) Bloch function
$u_{i}({\bf x})$ reduces to a constant, resulting in $B_{ij}({\bf G})=\delta_{{\bf G},0}$. This yields
\begin{equation}
{\bf F}_{w}=\frac{\pi e^2}{\hbar S^{2}}\sum_{i,j}\left[\delta f_{i}-\delta f_{j}\right]\left({\bf k}_{i}-{\bf k}_{j}\right)\left|U_{{\bf R}}({\bf k}_{i}-{\bf k}_{j})\right|^{2}\delta(\omega_{ij}),\label{eq:wind_free}
\end{equation}
which is essentially a Fermi golden rule expression for the electron
wind, where the rate of scattering, $\sim\left|U_{{\bf R}}({\bf k}_{i}-{\bf k}_{j})\right|^{2}$,
is weighted with the density of available final and initial states,
$\delta f_{i}-\delta f_{j}$, and the momentum transfer, ${\bf k}_{i}-{\bf k}_{j}$.
It has to be emphasized here that ${\bf k}_{i}$ becomes the true
momentum here, and not just a quasimomentum.

An example where the difference between momentum and quasimomentum
is emphasized can be constructed as follows. We still consider free
electrons, but unlike the previous example, we introduce a fictitious
periodic lattice (lattice constant $\sim a$) with an infinitesimal
periodic potential. If the unit cell is large compared to the inverse
Fermi momentum of free electrons (i.e., $a\gg1/q_{F}$), then a quasimomentum
of an electronic state near the Fermi surface, taken in the first
Brillouin zone, is much smaller in magnitude than the actual momentum
of the state. This is because a quasimomentum in the first Brillouin
zone cannot exceed $\sim1/a$. However, since the problem still concerns
free electrons (albeit in a different representation), Eq.~(\ref{eq:Fw_general})
should still yield momentum transfer prefactors, ${\bf k}_{i}-{\bf k}_{j}-{\bf G}$,
with the magnitude equal to the true Fermi momentum of free electrons.
This is realized in Eq.~(\ref{eq:Fw_general}) by appearance of non-vanishing
contributions with ${\bf G}\ne0$ . Specifically, a rapidly oscillating
free-electron wavefunction with momentum ${\bf p}$, being treated
as a Bloch wave in the lattice with the lattice constant $a$, is
represented as $e^{i{\bf p}{\bf x}}=e^{i{\bf k}{\bf x}}e^{i{\bf G}{\bf x}}$
in the first Brillouin zone of the fictitious lattice, where ${\bf p}={\bf k}+{\bf G}$,
and ${\bf G}$ is chosen so that $k\lesssim1/a\ll p$. The second
r.h.s. factor in the wavefunction representation is thus a rapidly
oscillating Bloch function ($G\sim p$), which, by virtue of Eq.~(\ref{eq:B_exp}),
yields a momentum transfer prefactor of a required magnitude in Eq.~(\ref{eq:Fw_general}).

Another example concerns an adsorbate-substrate system where the actual
substrate lattice (i.e., with an underlying periodic potential) is
present with the characteristic lattice constant of $a$. We assume
that the following conditions are satisfied: (i) the scattering potential
is smooth over the unit cell, i.e., $U_{{\bf R}}(k\sim1/a)\approx0$,
and (ii) the quisimomenta of electron states belonging to the Fermi
surface is much less in magnitude than $1/a$ (i.e., the Fermi surface
is small and centered around the $\Gamma$-point). Then, Eq.~(\ref{eq:Fw_general})
reduces to
\begin{align}
{\bf F}_{w}&=\frac{\pi e^2}{\hbar S^{2}}\sum_{i,j}\left[\delta f_{i}-\delta f_{j}\right]\left({\bf k}_{i}-{\bf k}_{j}\right)\nonumber\\
&\times\left|U_{{\bf R}}({\bf k}_{i}-{\bf k}_{j})B_{ij}\right|^{2}\delta(\omega_{ij}),\label{eq:Fw_general-1}
\end{align}
where $B_{ij}=B_{ij}({\bf G}=0)=s^{-1}\int_{s}d{\bf x}\, u_{i}^{*}({\bf x})u_{j}({\bf x}).$
This expression for the wind force is very similar to the one for
free electrons, Eq.~(\ref{eq:wind_free}), except that the scattering
amplitude now depends not only on the Fourier transform of the
adsorbate potential, $U_{{\bf R}}({\bf k}_{i}-{\bf k}_{j})$, but
also on the expansion coefficients of Bloch functions at $\Gamma$-point,
$B_{ij}$. Thus, if the two conditions specified above are satisfied,
then momentum transfer can be equated with the change in quasimomentum
with the only difference from the free electron picture being the
renormalization of the scattering potential by the band structure
effects via $B_{ij}$.

The last important scenario concerns essentially the previous example
with the only modification that the Fermi surface is centered not
around the $\Gamma$-point, but around some other high-symmetry point
located at the edge of the first Brillouin zone. Then, of course,
Eq.~(\ref{eq:Fw_general}) still applies, but significant fraction
of scattering events would result in the Umklapp scattering, i.e.,
${\bf k}_{i}-{\bf k}_{j}$ falls off the first Brillouin zone, but,
e.g., ${\bf k}_{i}-{\bf k}_{j}-{\bf G}$ is small compared to $1/a$.
However, instead of applying the general formalism, one can use a
more physically transparent approach. Specifically, we recall that the reciprocal lattice is periodic in a sense that quasimomentum ${\bf k}_{i}+{\bf G}$ (${\bf G}$ being an arbitrary translational
vector of the reciprocal lattice) is physically equivalent to ${\bf k}_{i}$. The standard convention is to center the
first Brillouin zone around the $\Gamma$-point, but this is not the
only possibility. In principle, on can center the first Brillouin
zone around an arbitrary point of the reciprocal lattice and then
measure a quasimomentum relative to that point. This ``translation''
of the Brillouin zone is often used when the ${\bf k}\cdot{\bf p}$
approach is applied to describe an electronic structure of various
materials with translational symmetry, e.g., lead salts (PbSe, PbS)\cite{Mitchell1966-581,Dimmock1971,Kang1997-1632}
or graphene.\cite{Semenoff1984-2449,CastroNeto2009-109} The only
pitfall of this approach is that Bloch functions are not periodic
on the lattice anymore, since they have to absorb the momentum change
between the new center of the Brillouin zone and the $\Gamma$-point.
However, this momentum change often cancels out in practical calculations
of various matrix elements of Bloch functions, e.g., in Eq.~(\ref{eq:B_exp}),
and, thus, even though the Brillouin zone is not centered around the
$\Gamma$-point any more, Eq.~(\ref{eq:Fw_general-1}) can still
be used if (i) the scattering potential is smooth, i.e., $U_{{\bf R}}(k\sim1/a)\approx0$,
and (ii) the Brillouin zone can be centered around the Fermi surface
so that quasimomenta laying on this surface have the magnitude $q_{F}\ll1/a$. 

Considering only scattering within a single band, so that an electron
state can be unambiguously specified by a quasimomentum alone, one
can substitute the summation over all the possible states above with
integration over quasimomentum within the first Brillouin zone
\begin{align}
{\bf F}_{w}&=\frac{\pi e^2\hbar^{-1}}{(2\pi)^{4}}\int d{\bf k}\int d{\bf k}'\,\left[\delta f({\bf k})-\delta f({\bf k}')\right]\nonumber\\
&\times\left({\bf k}-{\bf k}'\right)\left|U({\bf k}-{\bf k}')B_{{\bf k}{\bf k}'}\right|^{2}\delta(\omega_{{\bf k}{\bf k}'}).\label{eq:windf},
\end{align}
where $\omega_{{\bf k}{\bf k}'}=\omega_{\bf k}-\omega_{{\bf k}'}$. This expression for the wind force is the one that will be used to
calculate the electron wind force acting on covalently-bound adsorbates
on graphene. The next section is devoted to evaluation of scattering
potential and $B_{{\bf k}{\bf k}'}$-factors for graphene, as well as to justification
of assumptions leading to the derivation of Eq.~(\ref{eq:windf}).

\section{Wind force of a covalently-bound adsorbate on graphene}\label{sec:graphene}

The previous section resulted in the derivation of general expression
for the electron wind force upon an adsorbate on a substrate. We apply
these results to a specific case of covalently-bound adsorbates on
graphene.

\subsection{Electronic structure of graphene: Dirac fermion approximation\label{sub:Dferm}}

Low energy excitation as well as transport properties of doped graphene
can be described within the linearized tight-binding (or Dirac fermion)
approximation. The tight-binding Hamiltonian for graphene without
adsorbates is given by\cite{CastroNeto2009-109}
\begin{equation}
\hat{H}=-t\sum_{k,l}\hat{c}_{k}^{\dagger}\hat{c}_{l},\label{eq:tb_0-1}
\end{equation}
where operator $\hat{c}_{k}^{\dagger}$ ($\hat{c}_{k}$) creates (destroys)
an electron on a $p_{z}$ orbital of carbon at position $i$ within
the graphene lattice. The summation is performed only over pairs of
indices corresponding to nearest-neighbor sites. The hopping integral
is taken to be $t=2.7$ eV. We choose the orientation of the graphene
lattice so that two carbon atoms within a unit cell are connected
by vector $\bm{\delta}_{1}=a(0,1)$, where $a=1.42\,{\rm \AA}$ is
the carbon-carbon distance in the pristine graphene. Other two carbon-carbon
vectors are $\bm{\delta}_{2}=a(-\frac{\sqrt{3}}{2},-\frac{1}{2})$
and $\bm{\delta}_{3}=a(\frac{\sqrt{3}}{2},-\frac{1}{2})$.
\begin{figure}
\includegraphics[width=3in]{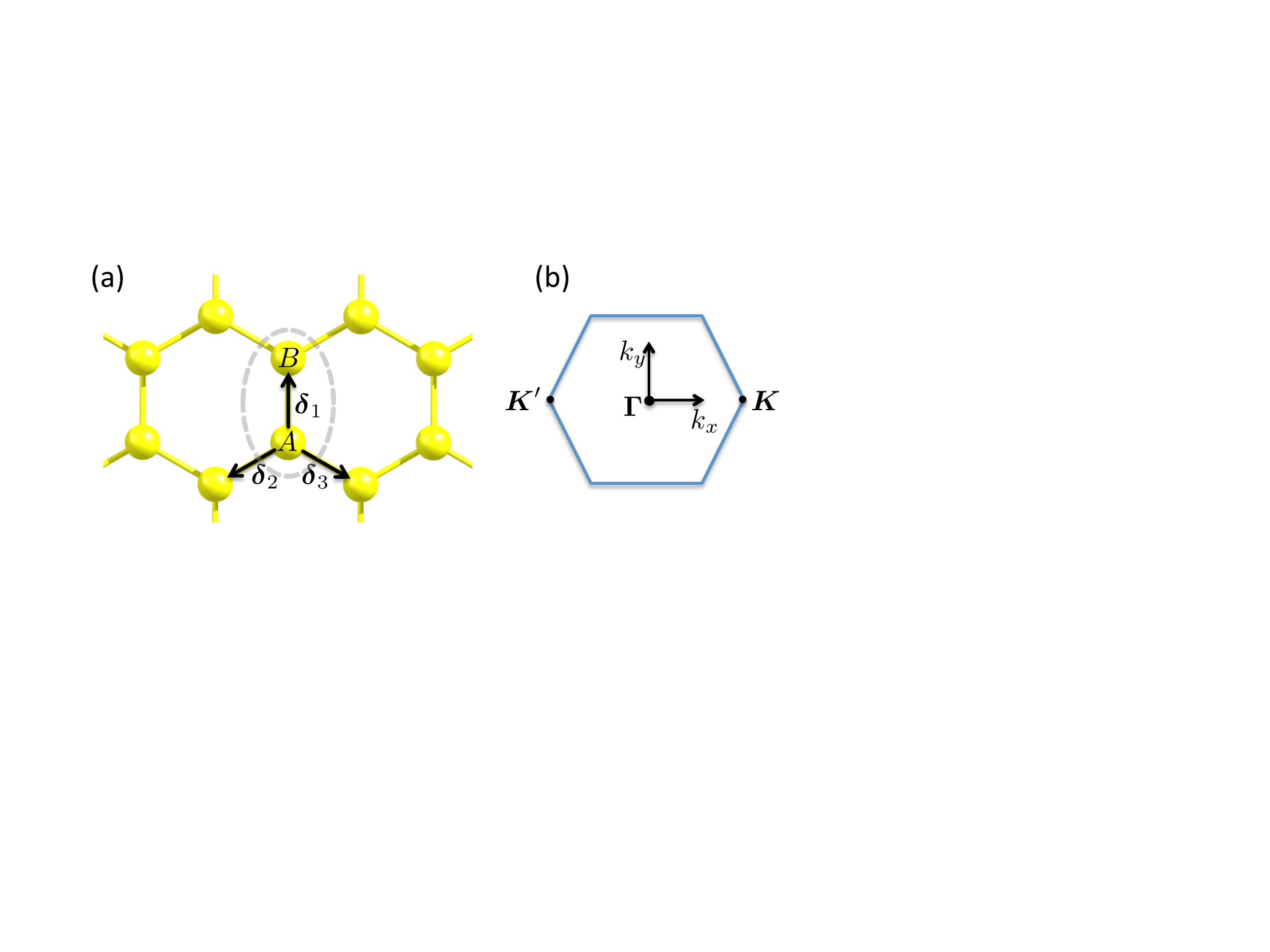}
\caption{\label{fig:real_reciprocal}(a) Portion of graphene honeycomb lattice. (b) The first Brillouin zone of the reciprocal lattice of graphene. }
\end{figure}
This choice of vectors is illustrated in Fig.~\ref{fig:real_reciprocal}(a).
Accordingly, the Brillouin zone of graphene is the hexagon with two
sides parallel to $k_{x}$ axis, and the two inequivalent Dirac points
are at ${\bf K}=\frac{4\pi}{3\sqrt{3}a}(1,0)$ and ${\bf K}'=-{\bf K}.$
These two high-symmetry points, as will as the $\bm{\Gamma}$-point,
are shown in Fig.~\ref{fig:real_reciprocal}(b). Seeking eigenfunctions
of the Hamiltonian operator in Eq.~(\ref{eq:tb_0-1}) as Bloch waves
results in a Schrodinger-like equation for the Bloch functions, $\hat{H}_{{\bf k}}|u\rangle=E|u\rangle$,
where $|u\rangle=(u_{A},u_{B})^{T}$ is the spinor representation
of a Bloch function, $u_{A}$ ($u_{B}$) being an amplitude to find
an electron on the $p_{z}$ orbital of the first (second) atom of
the graphene unit cell. These atoms are marked in Fig.~\ref{fig:real_reciprocal}(a)
by sublattice symbols $A$ and $B$ within a dashed grey oval. Operator
$\hat{H}_{k}$ reads as 
\begin{equation}
\hat{H}_{{\bf k}}=\begin{pmatrix}0 & \Delta({\bf k})\\
\Delta^{*}({\bf k}) & 0
\end{pmatrix},
\end{equation}
where $\Delta({\bf k})=-t\sum_{i=1}^{3}e^{i(\bm{\delta}_{i}\cdot{\bf k})}$.
At ${\bf k}={\bf K}$ and ${\bf k}={\bf K}'$, $\Delta({\bf k})$
vanishes exactly, and the linear expansion around these points yields
\begin{equation}
\Delta({\bf k}+{\bf K})\approx\frac{3}{2}ta(k_{x}-ik_{y})=vke^{-i\theta_{{\bf k}}},
\end{equation}
and
\begin{equation}
\Delta({\bf k}+{\bf K}')\approx\frac{3}{2}ta(-k_{x}-ik_{y})=-vke^{i\theta_{{\bf k}}},
\end{equation}
where $\theta_{{\bf k}}$ is the angle between vector ${\bf k}$ and
the $k_{x}$ axis and $v=\frac{3}{2}ta$ is the Fermi velocity of
electrons in graphene. As a result, the linearized version of the
$H_{{\bf k}}$ Hamiltonian in the neighborhood of ${\bf K}$ and ${\bf K}'$
points can be written as, respectively,
\begin{equation}
H_{{\bf K}}({\bf k})=v(\hat{\bm{\sigma}}\cdot{\bf k}),\label{eq:HK}
\end{equation}
and 
\begin{equation}
H_{{\bf K}'}({\bf k})=v(\hat{\bm{\sigma}}'\cdot{\bf k}),\label{eq:HKp}
\end{equation}
where $\hat{\bm{\sigma}}=(\hat{\sigma}_{x},\hat{\sigma}_{y})$ and
$\hat{\bm{\sigma}}'=(-\hat{\sigma}_{x},\hat{\sigma}_{y})$. In the
spirit of the last example discussed in Sec.~\ref{sub:wind_force},
quasimomentum ${\bf k}$ in Eqs.~(\ref{eq:HK}) and (\ref{eq:HKp})
is measured relative to ${\bf K}$ and ${\bf K}'$ points of the Brillouin
zone, respectively. Eigenenergies of these Hamiltonians are given
by $E=\pm vk$. Electronic states with positive (negative) energies
are said to belong to the $\pi^{*}$($\pi$) band. Therefore, the
electronic structure of graphene at not very high energies is said
to be given by two Dirac cones, centered at ${\bf K}$ and ${\bf K}'$.
At these conditions, each electronic state can be unambiguously specified
by multi-index $i=({\bf k}_{i},b_{i},c_{i})$, where ${\bf k}_{i}$
is the quasimomentum, $b_{i}=+1(-1)$ if the state belongs to the
$\pi^{*}$($\pi$) band, and $c_{i}={\bf K}$ or ${\bf K}'$ to specify
the Dirac cone. The spin variable is omitted. Eigenfunctions of these
two linearized Hamiltonian operators, i.e., Bloch functions, are found
to be
\begin{equation}
|u_{i}\rangle=\begin{cases}
\left(1,\, b_{i}e^{i\theta_{i}}\right)^{T}/\sqrt{2}, & {\rm if}\, c_{i}={\bf K}\\
\left(-1,\, b_{i}e^{-i\theta_{i}}\right)^{T}/\sqrt{2}, & {\rm if}\, c_{i}={\bf K}'
\end{cases}\label{eq:Bloch}
\end{equation}
where $\theta_{i}=\theta_{{\bf k}_{i}}$. This spinor representation
of Bloch functions becomes very convenient for calculation of various
matrix elements. For example, evaluation of the overlap of Bloch functions
of two electron states within the same ${\bf K}$-cone reduces to
a simple dot-product yielding
\begin{align}
B_{ij}^{{\bf K}}&=B_{ij}^{{\bf K}}({\bf G}=0)=\langle u_{b_{i},{\bf K}}({\bf k}_{i})|u_{b_{j},{\bf K}}({\bf k}_{j})\rangle\nonumber\\
&=\frac{1+b_{i}b_{j}e^{-i\theta_{ij}}}{2},\label{eq:over1}
\end{align}
where $\theta_{ij}=\theta_{i}-\theta_{j}=\theta_{{\bf k}_{i}}-\theta_{{\bf k}_{j}}$.
Similarly evaluating $B_{ij}^{{\bf K}'}$, one obtains $|B_{ij}^{{\bf K}}|^{2}=|B_{ij}^{{\bf K}'}|^{2}=\cos^{2}(\theta_{ij}/2)$
or $\sin^{2}(\theta_{ij}/2)$ for electronic states $i$ and $j$
belonging to the same (e.g., $\pi^{*}$) or different bands, respectively,
within a single Dirac cone.

Another important identity which can be obtained from Eq.~(\ref{eq:Bloch})
is 
\begin{equation}
\int_{0}^{2\pi}d\theta_{i}\,|u_{i}\rangle\langle u_{i}|=\pi\hat{\sigma}_{0},\label{eq:uu_aver}
\end{equation}
where $\hat{\sigma}_{0}$ is the two-by-two identity matrix.

\subsection{Effective potential of adsorbate\label{sub:effective_potential}}

The wind force acting upon an adsorbate on graphene is caused by scattering
of itinerant electrons in graphene by an effective potential of the
adsorbate, Eq.~(\ref{eq:Apot}). In the case of adsorbates covalently-bound
to graphene, this potential is expected to have two distinct contributions,\cite{Solenov2012-095504}
$U=U_{C}+U_{d}$, where the first one, $U_{C}$, is the Coulomb potential
induced by the charge localized on the adsorbate. The $U_C$ contribution
is significant if an adsorbate is sufficiently electronegative (electropositive)
so it becomes negatively (positively) charged upon the attachment
to graphene. The second contribution, $U_{d}$, is due to a defect
formed within the electronic structure of graphene by the attachment
of the adsorbate. Specifically, a covalently-bound adsorbate modifies
the electronic structure of graphene by changing the hybridization
of a certain number of carbon atoms in graphene. For example, a single
atom of oxygen, attached to graphene in the epoxy configuration [see Fig.~\ref{fig:Schematic}(a,c)], changes the hybridization of two carbon atoms it is attached to from $sp^{2}$ to $sp^{3}$. An $sp^{3}$-hybridized carbon atom
does not have a non-hybridized $p_{z}$ orbital to participate in
graphene delocalized $\pi$ (valence) and $\pi^{*}$ (conduction)
bands. This breaks the translational invariance of the system and,
therefore, forms an electron-scattering defect. As will become clear
shortly, the Coulomb contribution is dominant for adsorbates considered
in this paper so we discuss the Coulomb contribution first.

The bare (i.e., unscreened) Coulomb potential generated by a charged
adsorbate within a graphene sheet is given by
\begin{equation}
U_{C}^{0}(r)=\frac{Ze}{\sqrt{r^{2}+h^{2}}},
\end{equation}
where $h$ and $r$ are the effective distance of the adsorbate from
the graphene sheet and the distance within the graphene sheet, respectively.
The effective charge of the adsorbate in atomic units is denoted by
$Z$. The in-plane Fourier transform of this potential is
\begin{equation}
U_{C}^{0}(q)=\int d{\bf r}\, U_{C}^{0}(r)e^{-i{\bf q}{\bf r}}=\frac{2\pi Ze}{q}e^{-qh}\approx\frac{2\pi Ze}{q},\label{eq:Cbare}
\end{equation}
where the last approximate equality is, in fact, very accurate since
for scattering near the Fermi surface of back-gated graphene one has
$q\sim k_{F}$ resulting in $qh\ll1$ for all realistic graphene samples
($h\sim1{\rm \AA}$ for small adsorbates).

Once the bare Coulomb potential of the adsorbate, Eq.~(\ref{eq:Cbare}),
is present in a realistic graphene sample, it becomes screened by
the electrostatic polarization of doped graphene and environment.
Within a perturbative linear response approximation, the \emph{screened} Coulomb potential is given by \cite{DasSarma2011-407,Solenov2012-095504}
\begin{equation}
U_{C}(q)=\frac{2\pi Ze}{\kappa(q+q_{TF})},\label{eq:Cscreen}
\end{equation}
where $q_{TF}=4q_{F}e^2/(\kappa\hbar v_{F})\approx9q_{F}/\kappa$ is
the Thomas-Fermi screening wavevector.\cite{CastroNeto2009-109}
The effective dielectric constant of environment is given by $\kappa=(\kappa_{1}+\kappa_{2})/2$
for a graphene sheet sandwiched between two substrates with dielectric
constants $\kappa_{1}$ and $\kappa_{2}$.\cite{Ponomarenko2009-206603}
In what follows, a SiO$_{2}$ substrate in vacuum is assumed so $\kappa=(1+\kappa_{{\rm SiO_{2}}})/2=2.5$,
and, therefore, $q_{TF}\approx3.6q_{F}$.

Eq.~(\ref{eq:Cscreen}) is only valid in the perturbative regime when $Z$ is not very large. Specifically, non-perturbative regime can only be realized when $Z\gtrsim 1$.\cite{Shytov2007-236801,Biswas2007-205122,Shytov2007-246802,Terekhov2008-076803} It will be shown in Sec.~\ref{sec:res} that $Z$ is sufficiently below this threshold for the adsorbates considered in this paper. Eq.~(\ref{eq:Cscreen}) is expected to be accurate at these conditions. 

The qualitative difference between unscreened and screened forms of
the Coulomb potential above is that the former, $U_{C}^{0}(q)$, is
inherently a long-range potential, diverging at $q\rightarrow0$.
In contrast, the screened potential is short-range due to the presence
of the finite screening length, $\sim1/q_{TF}$. In fact, important
scattering events (i.e., those contributing to the wind force) occur
near Fermi surface, i.e., $q\sim q_{F}$. Since $q_{TF}$ is larger
than $q_{F}$ for not very polar substrates (e.g., ${\rm SiO_{2}}$),
the denominator in Eq.~(\ref{eq:Cscreen}) is dominated by $q_{TF}$,
and therefore, the screened Coulomb potential can be assumed \emph{contact}
(i.e., $q$-independent) 
\begin{equation}
U_{C}^{c}\approx\frac{2\pi Ze}{\kappa q_{TF}}=\frac{2\pi Ze}{4q_F}\frac{\hbar v_F}{e^2}\approx\frac{2\pi Ze}{9q_{F}},\label{eq:Ccontact}
\end{equation}
which will simplify calculations to follow. However, it has to be
noted here that $U_{C}$ can only be considered contact at quasimomenta
around the Fermi surface. The effective size of the actual potential
in Eq.~(\ref{eq:Cscreen}) is given by $1/q_{TF}$, so that at, for example, $E_{F}=0.2$ eV (easily reached in back-gated graphene
samples) one has $q_{F}a\approx0.05$, and, therefore, $q_{TF}a\approx0.18$.
Therefore, the size of the potential is still significantly larger
than the size of the graphene unit cell. Hence, even though the contact
form of the screened Coulomb potential, Eq.~(\ref{eq:Ccontact}),
is accurate at $q\sim q_{F}$, scattering at $q\sim1/a\gg q_{TF}$
is suppressed {[}see Eq.~(\ref{eq:Cscreen}){]}, so that the scattering
between the Dirac cones can be safely neglected. This justifies the
approximation of scattering within a single band (i.e., Dirac cone
here) assumed in Eq.~(\ref{eq:windf}). Note also that Eq.~(\ref{eq:Ccontact}) does not depend on effective dielectric constant $\kappa$. Both, Coulomb interaction between conduction electrons and the adsorbate potential exist in the same dielectric environment. As a result, $\kappa$ in Eq.~(\ref{eq:Ccontact}) is canceled.

The second contribution to the effective potential of an adsorbate
comes from the formation of a defect. The effective ``size'' of
this potential - exclusion of carbon non-hybridized $p_{z}$ orbitals
from the delocalized $\pi$-system - is expected to be on the order
of $a$. The only energy parameter in the system is the hopping integral
of graphene in the tight-binding representation, $t$. Thus, one can
estimate the amplitude of the unscreened defect potential from the
dimensional analysis as $U_{d}^{0}\sim ta^{2}/e$. The more detailed
analysis provided in App.~\ref{sec:point_def} corroborates this
estimate. The ratio of two contributions to the scattering potentials
is (at $q=q_{F}$)
\begin{equation}
U_{d}^{0}/U_{C}^{0}\approx\frac{0.028}{Z}\left(\frac{E_{F}}{t}\right),
\label{eq:Ud0}
\end{equation}
where we assume $t=2.7$ eV. Since the doping level of graphene is
typically not higher than $\sim$0.5 eV and the charge of an adsorbate
could be $|Z|\approx0.2-1$ for the adsorbates described in this work
(see Sec.~\ref{subsec:Hopping}), the contribution of the defect potential
is much smaller than that of the Coulomb interaction. The actual defect
potential can lead to exclusion of several $p_{z}$ orbitals, e.g.,
two for oxygen in the equilibrium state, see Fig.~\ref{fig:Schematic}(d).
Furthermore, the defect potential Eq.~(\ref{eq:Ud0}) is not screened by the substrate.
However, these additional factors do not change the qualitative
conclusion drawn: the defect contribution to the electron wind force
is small compared to that of Coulomb interaction and will be neglected
henceforth.

Note that this conclusion is based on the fact that the wind force originates from scattering near the Fermi surface. This is not necessarily the case for other observables. For example, static interaction between two adsorbates involves scattering at all energies and not just around the Fermi level,\cite{Solenov2013-115502} and, thus, Eq.~(\ref{eq:Ud0}) is not  applicable in that case.

\subsection{Electron wind force}

It has been established in the previous subsection that (i) the electron
scattering in graphene at the Fermi level is dominated by the Coulomb
potential of a charged adsorbate, and (ii) the scattering by this
potential at momenta comparable with $1/a$ is strongly suppressed.
Under these conditions, Eq.~(\ref{eq:windf}) can be used to evaluate
the electron wind force in a doped graphene. The deviation of the
single-electron population from the equilibrium Fermi-Dirac one in
the non-equilibrium steady-state case when the constant current density
${\bf j}$ is present in the graphene is given in the relaxation time
approximation as\cite{Ashcroft1976}
\begin{equation}
\delta f({\bf k})=-\frac{4\pi({\bf j}\cdot{\bf k})}{ev_{F}q_{F}^{2}}\delta(k-q_{F}),\label{eq:FDcurrent}
\end{equation}
Indeed, the current density carried by an electron in a state with momentum
${\bf k}$ is ${\bf j}_{{\bf k}}=-S^{-1}ev_F\hat{\bf k}$, where  $\hat{\bf k}$ is the unit vector in the direction of quasimomentum
${\bf k}$, and the minus sign is due to the negative charge of electron.
The total current density is then given by
\begin{equation}
\sum_{{\bf k}}{\bf j}_{{\bf k}}f({\bf k})=\frac{S}{(2\pi)^{2}}\int d{\bf k}\,{\bf j}_{{\bf k}}\delta f({\bf k}),
\end{equation}
which produces exactly ${\bf j}$.

Eq.~(\ref{eq:windf}) describes the elastic scattering (due to the
factor of $\delta(\omega_{{\bf k}{\bf k}'})$) exactly at the Fermi
level due to delta-function in Eq.~(\ref{eq:FDcurrent}), and, therefore,
the scattering is \emph{intraband}. At this condition, the Bloch function
overlap is given by $B_{{\bf k}{\bf k}'}=\cos^{2}\left(\theta_{{\bf k}{\bf k}'}/2\right)$
(see Sec.~\ref{sub:Dferm}). Substituting this expression along with
Eqs.~(\ref{eq:FDcurrent}) and (\ref{eq:Ccontact}) into Eq.~(\ref{eq:windf})
results in 
\begin{align}
{\bf F}_w&=-\frac{\hbar v_F Z^2}{16eq^4_F}\int d{\bf k}d{\bf k}'\:({\bf k}-{\bf k}')\cos^2(\theta_{{\bf k}{\bf k}'}/2)\delta(\omega_{kk'})\nonumber\\
&\times\left[({\bf j}\cdot{\bf k})\delta(k-q_F)-({\bf j}\cdot{\bf k}')\delta(k'-q_F)\right].
\end{align}
Using equality $\delta(\omega_{kk'})=v^{-1}_F\delta(k-k')$ and integrating over the radial components of ${\bf k}$ and ${\bf k}'$ one obtains
\begin{equation}
{\bf F}_w=-\frac{\hbar Z^2}{16e}\int d\theta_{\bf k} d\theta_{{\bf k}'}\:(\hat{\bf k}-\hat{\bf k}')\cos^2(\theta_{{\bf k}{\bf k}'}/2)
({\bf j}\cdot\hat{\bf k}-{\bf j}\cdot\hat{\bf k}')
\end{equation}
Only the vector component parallel to ${\bf j}$ survives the integration resulting in
\begin{equation}
{\bf F}_w=-\frac{\hbar Z^2{\bf j}}{16e}\int d\theta d\theta'\:\left(\cos\theta-\cos\theta'\right)^2\cos[(\theta_1-\theta_2)/2]
\end{equation}
This integral is straightforwardly evaluated finally yielding a very compact expression
for the electron wind force
\begin{equation}
{\bf F}_{w}=-\frac{\hbar{\bf j}}{e}\left(\frac{\pi Z}{4}\right)^{2}.\label{eq:wforce}
\end{equation}
In the derivation of this expression we have not accounted for the
spin and Dirac cone (${\bf K}$ and ${\bf K}'$) degeneracies. However,
these degeneracies would enter Eq.~(\ref{eq:wforce}) only through
the current density. Therefore, if ${\bf j}$ is the \emph{total}
current density that already includes all the degeneracies, then
formally the same Eq.~(\ref{eq:wforce}) gives the total wind force
including the effect of all the degeneracies.

Eq.~(\ref{eq:wforce}) has been derived within the lowest order approximation
with respect to the adsorbate scattering potential, $U$. This approximation
is expected to be accurate if the effective charge of an adsorbate,
$Z$, is not very large. As will be seen shortly, $|Z|\approx0.2-1$
for both oxygen and nitrogen, and, therefore, the ratio of the screened
Coulomb potential energy to the Fermi energy is given by
\begin{equation}
\lambda=eU_{C}(q_{F})q_{F}^{2}/\hbar v_{F}q_{F}\approx|Z|/2\approx0.1-0.5.\label{eq:UC_unitless}
\end{equation}
This ratio is less than $1$ but not by much, thus raising a question on the
validity and accuracy of the weak scattering approximation. In other
words, is the first non-vanishing contribution to the wind force,
${\bf F}_{1,neq}^{-}$, in Eq.~(\ref{eq:tforce}) enough to obtain
an accurate result, or higher order corrections with respect to the
adsorbate potential (e.g., ${\bf F}_{2,neq}$) have to be included?
This problem is addressed in detail in App.~\ref{app:strong} with the conclusion that it is not $\lambda$ which has to be compared to
$1$, but rather $\lambda/2\pi\approx0.02-0.08$ for graphene-adsorbate
complexes considered in this work. As a result, a weak scattering
approximation is justified and accurate.

\section{Results and discussion}\label{sec:res}

In order to parameterize our general analytical results obtained above for specific functional groups ($-{\rm O}-$ and $-{\rm NH}-$) we have performed accurate \emph{ab initio} electronic
structure theory calculations using the VASP code.\cite{Kresse1996-15,Kresse1996-11169}
Specifically, we have used the density functional theory (DFT) parameterized
by the Perdew, Burke and Ernzerhof (PBE) functional\cite{Perdew1996-3865,Perdew1997-1396}
and projector-augmented wave (PAW) potentials.\cite{Blochl1994-17953,Kresse1999-1758}
All the calculations have been performed using $5\times5$ graphene supercell (50 carbon atoms). 

Due to periodic boundary conditions in all three directions, the actual
geometry of the system being calculated is a 3D layered structure
with multiple graphene sheets placed periodically along the $z$-axis.
On one hand, the distance between the sheets has to be as large as
possible to avoid interaction between them. On the other hand,
very large distance makes calculations computationally expensive.
Furthermore, it has been shown by Topsakal et. al.~\cite{Topsakal2013-5943}
that there is an optimal distance between the sheets in the case of
charged graphene. Specifically, 20 \AA{} of vacuum between the sheets
allows (i) to avoid significant interactions between the sheets and
also (ii) to guarantee the minimal effect of electron spilling into
vacuum.\cite{Topsakal2013-5943} This optimal distance has been adopted in
all our calculations. Monopole and dipole corrections were included
to improve convergence with respect to the size of the cell.\cite{Neugebauer1992-16067,Makov1995-4014,Suarez2011-146802}
Energy difference of $10^{-5}$ eV was selected as a minimum value
for electronic relaxations and $10^{-2}$ eV/\AA{} was selected for
ionic relaxations. The Brillouin zone was sampled with the $5\times5\times1$
Monkhorst-Pack grid.\cite{Monkhorst1976-5188} 

\subsection{Energy and charge along hopping trajectory}
\label{subsec:Hopping}

\begin{figure}
\includegraphics[width=2.5in]{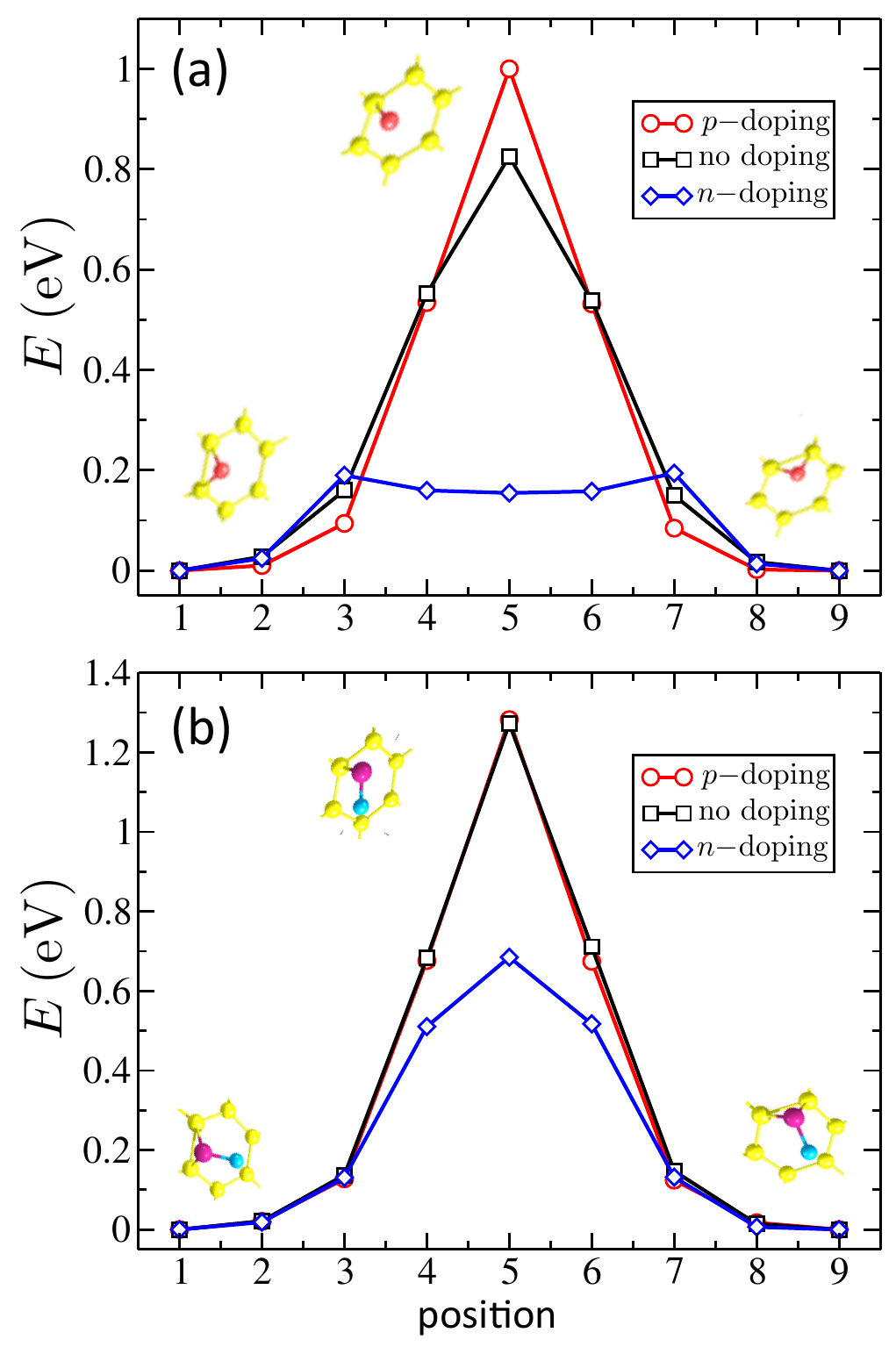}
\caption{\label{fig:Eprof} Energy profile along the hopping trajectory, measured relative to the energy of the equilibrium configuration. Positions 1 and 9 correspond to equilibrium configurations, position 5 corresponds to the transition state. (a) Hopping trajectory of epoxy group at various degrees of graphene
doping. (b) Energy profile along the hopping
trajectory for amine group at various levels of doping. }
\end{figure}

\begin{figure}
\includegraphics[width=3in]{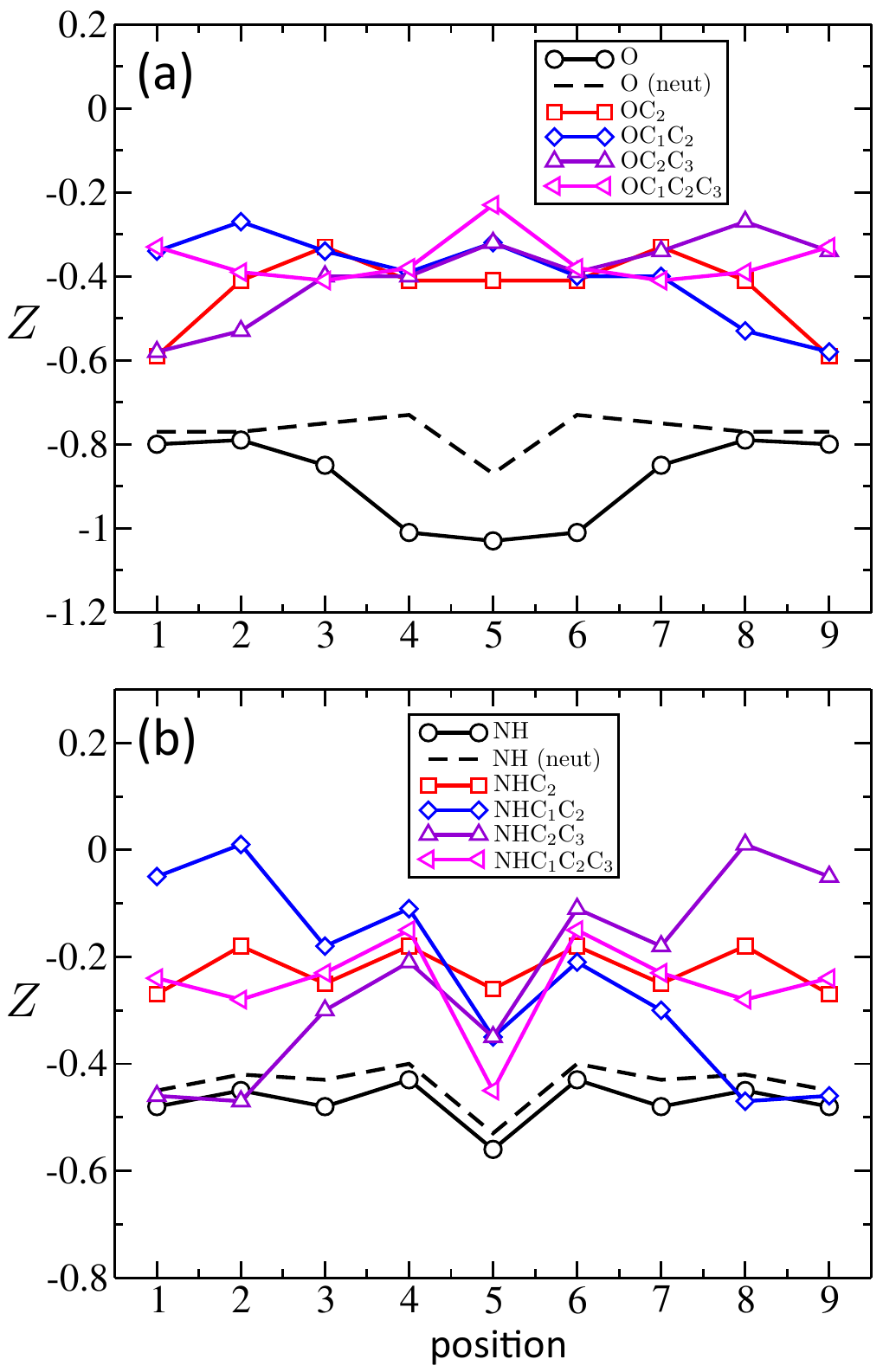}
\caption{\label{fig:charges} Partial charges for epoxy (a) and amine (b) groups on graphene ($5\times5$ cell, 50 carbon atoms). Solid lines in both panels correspond to negatively doped graphene ($n=-7.63\times10^{13}~cm^{-2}$). The dashed lines show the partial charge on the adsorbate (epoxy or amine group) for the case of undoped graphene. $\rm C_1$ and $\rm C_2$ are the carbon atoms an adsorbate is attached to in the initial position. $\rm C_2$ and $\rm C_3$ are the atoms the adsorbate is attached to in the final position. The covalent bond to $\rm C_2$ is never broken.}
\end{figure}

The diffusion of epoxy and amino groups along the surface of graphene
proceeds via a transition state where one of the two ${\rm O-C}$ or ${\rm N-C}$ bonds
is broken, Fig.~\ref{fig:Schematic}(b). Energy profiles for these groups, defined as the variation of energy of the adsorbate-graphene system measured relative to the energy of the equilibrium configuration,  have been calculated using the nudged elastic band method.\cite{Sheppard2008-134106} A spring constant of 5.0
eV/\AA{}$^{2}$ was used in these calculations. The energy profiles for epoxy and amino groups for various levels of graphene doping are shown in Fig.~\ref{fig:Eprof}(a) and (b), respectively.
As shown in Tab.~\ref{tab:TS} and Fig.~\ref{fig:Eprof}, the charge
doping either increase or decrease the energy barrier by removing
or adding an electron, in good agreement with results reported before.\cite{Suarez2011-146802,Solenov2012-095504} 
\begin{center}
\begin{table*}
\caption{\label{tab:TS}Activation energies and attempt frequencies for diffusion
of epoxy (${\rm O}$) and amine (${\rm NH}$) groups on graphene for
various doping levels ($5\times5$ supercell)}
\begin{tabular}{|l|l|l|l|l|l|l|}
\hline 
System  & $n$ ($cm^{-2}$) & $E_{F}$ (eV) & $E_{a}^{{\rm O}}$ ($eV$) & $\nu^{{\rm O}}$ (THz) & $E_{a}^{{\rm NH}}$ ($eV$) & $\nu^{{\rm NH}}$ (THz)\tabularnewline
\hline 
Positive doping (+1)  & $+7.63\times10^{13}$ & $-0.9$ & $1.0$ & 51 & 1.28 & 78\tabularnewline
\hline 
Neutral  & $0$ & $0$ & $0.82$ & 30 & 1.27 & 90 \tabularnewline
\hline 
Negative doping (-1)  & $-7.63\times10^{13}$ & $+0.9$ & $0.15$ & 14 & 0.68 & 72\tabularnewline
\hline 
\end{tabular}
\end{table*}
\par
\end{center}
Partial charge analysis of atoms related to electromigration of functional groups is required
to assess the strength of the adsorbate Coulomb scattering potential,
Eq.~(\ref{eq:Ccontact}). It is however impossible to rigorously
define the charge of an adsorbate since there is no unique surface separating
an adsorbate from the rest of the system.\cite{Sorbello1997-159}
On the other hand, it is physically clear that since the electron
wind force stems from scattering of \emph{mobile} electrons in the
conduction band of doped graphene, only \emph{localized} charge has
to be associated with an adsorbate. We thus define the adsorbate
charge as an excess charge present on a functional group itself combined
with the charge sitting on the $sp^{3}$-hybridized carbon atoms of
graphene that the adsorbate is attached to. However, even this definition of the localized charge is not unambiguous since the entire concept of orbital hybridization
is imprecise and only qualitative even in an ideal situation of a
molecule in its optimal (equilibrium) geometry. In our case, we have to consider
the evolution of the hybridization along the hopping trajectory (Fig.~\ref{fig:Eprof}),
where bonds are formed and broken depending on a position along the
trajectory. To estimate the effect of this imprecision we analyze the charge not
only on a functional group itself, but also on all the graphene carbon atoms that the functional group attaches to when going along the hopping
trajectory from one equilibrium position to the other. Figs.~\ref{fig:charges}(a) and \ref{fig:charges}(b)
show the result of the Bader charge analysis\cite{Henkelman2006-254,Bader_scripts} for epoxy and amine groups on graphene, respectively. 
All the results are shown for the case of negatively-doped graphene ($n=-7.63\times10^{13}~{\rm cm}^{-2}$), except for those given by the dashed lines (undoped graphene).

As is seen in Fig.~\ref{fig:charges}(a), a significant portion of the localized charge is located on the carbon atoms an adsorbate is covalently bound to. However, the dependence of the partial charge on exactly which carbons atoms (out of $\rm C_1$, $\rm C_2$ and $\rm C_3$) are included into the definition of the localized charge is not too large and can be enclosed in a range $Z_{\rm epoxy}=-0.4\pm0.2$. Similar analysis can be performed for the amino group, Fig.~\ref{fig:charges}(b), yielding $Z_{\rm amino}=-0.25\pm0.1$. The variation of these charges with doping is within the assigned error bar.

We would like to emphasize here that the obtained error bar for $Z_{\rm epoxy}$ and $Z_{\rm amino}$ is not related to any inaccuracies in electronic structure theory calculations, which can be alleviated by more accurate calculations using, e.g., more advanced electronic structure theory methods. Instead, the presence of this error bar reflects a fundamental problem - impossibility to rigorously define the charge of a functional group when it is attached to a substrate. A reformulation of the electron wind force treating all the charge carriers (i.e., mobile and localized) on the same footing, would be required to evaluate the driving force of electromigraiton more accurately,\cite{Sorbello1997-159} which is beyond the scope of the current paper. 

Dashed lines in Fig.~\ref{fig:charges} show the partial charge of epoxy (panel a) and amino (panel b) groups in the case of the undoped graphene. The significant difference between the former partial charge with and without graphene doping gives an insight into why the negative doping of graphene results in a drastic decrease of the activation energy of oxygen diffusion. Due to a high electronegativity of oxygen, it typically forms two covalent bonds in compounds thus effectively completing its electronic shell up to neon. This is the case when an oxygen atom sits in its equilibrium configuration on graphene, Figs.~\ref{fig:Schematic}(a) and (c). However, in the transition state, Fig.~\ref{fig:Schematic}(b), one of these two bonds is broken and to still be able to complete its electronic shell oxygen has to ``pull" an electron from graphene. At negative doping, it is energetically easier for an epoxy group to pull an electron out of graphene. This (i) results in a stabilization of the transition state configuration by lowering its energy and (ii) makes a partial charge of oxygen approach $-1$, as is seen in Fig.~\ref{fig:charges}(a). In the undoped situation, a graphene lacks electrons it can easily donate to oxygen, resulting in the relatively high diffusion barrier and only a weak variation of oxygen partial charge over the hopping trajectory, dashed line in Fig.~\ref{fig:charges}(a).

Nitrogen is a much weaker electron acceptor compared to oxygen. This results in (i) a small difference between partial charges of amino group in the undoped and negatively dopes cases, Fig.~\ref{fig:charges}(b), and (ii) not as dramatic reduction of the diffusion barrier (compared to oxygen) with negative doping, Fig.~\ref{fig:Eprof}(b).

\subsection{Attempt Frequency}

The diffusion of the epoxy and amino groups on graphene can be analyzed by using the transition state theory (TST). The rate
of elementary hopping is obtained for the electronic structure
theory calculations via the formalism developed by Vineyard\cite{Vineyard1957-121}
\begin{equation}
\Gamma=\nu_{0}e^{-E_{a}/k_{B}T},\label{eq:diff_rate}
\end{equation}
where $\nu_{0}$ is the attempt frequency, calculated in the harmonic
approximation\cite{Vineyard1957-121} using the vibrational normal modes of
the equilibrium and transition states.\cite{VTST_scripts} The honeycomb
geometry of the graphene lattice and that an adsorbate can jump from its specific position to four adjacent ones can be accounted for yielding the diffusion coefficient as (detailed derivation is given in App.~\ref{sec:diff})
\begin{equation}
D=\frac{3}{4}a^{2}\Gamma=d^{2}\Gamma,\label{eq:diffc}
\end{equation}
where $d=\sqrt{3}a/2\approx1.23\,{\rm \AA}$ is the distance corresponding
to the elementary hopping event and $a\approx 1.42\,{\rm \AA}$ is the $\rm C-C$ bond length. The attempt frequency calculated for epoxy and amino groups at various levels of graphene doping is given in Table~\ref{tab:TS}.

Eqs.~(\ref{eq:diff_rate}) and (\ref{eq:diffc}) are obtained within the TST approximation.\cite{Haenggi1990-251} Deviations from this approximation are possible if the environment-induced friction suppressing the motion of an adsorbate is too high or too low. However, it is plausible to assume that this friction is not too high at realistic conditions. Indeed, if it was too high so that the vibration of an epoxy group on graphene was {\em  overdamped}, then the corresponding infrared (IR) absorption lines would not be resolved. However, these lines are typically resolved in experimental IR spectra\cite{Park2009-15801,Zhang2010-1112} so the friction can only be weak to moderate. At these conditions, TST estimates the diffusion rate from below.\cite{Dieterich1977-177} Therefore, the actual diffusion coefficient of adsorbates on graphene is expected to be no lower than the one given by Eqs.~(\ref{eq:diff_rate}) and (\ref{eq:diffc}).  

\subsection{Electromigration: Numerical results}

The total driving force of electromigration for the charged adsorbate
on graphene is given by\cite{Sorbello1997-159,Solenov2012-095504}
\begin{equation}
{\bf F}={\bf F}_{d}+{\bf F}_{w}=eZ{\bf j}/\sigma-\frac{\hbar{\bf j}}{e}\left(\frac{\pi Z}{4}\right)^{2}=eZ^{*}{\bf j}/\sigma,\label{eq:fforce}
\end{equation}
where the\textit{ effective} charge of the adsorbate is introduced
as $Z^{*}=Z-\frac{\sigma}{8\sigma_{0}}Z^{2}$ and $\sigma_{0}=\frac{2e^{2}}{\hbar\pi^{2}}$
is the so-called universal conductivity of graphene.\cite{CastroNeto2009-109} The second contribution, the electron wind force, is given by Eq.~(\ref{eq:wforce}). The first contribution, {\em the direct force}, originates from the direct action of the current-driving external electric field on the charge of the localized charge associated with the functional group.
 
\begin{figure}
\includegraphics[width=2.5in]{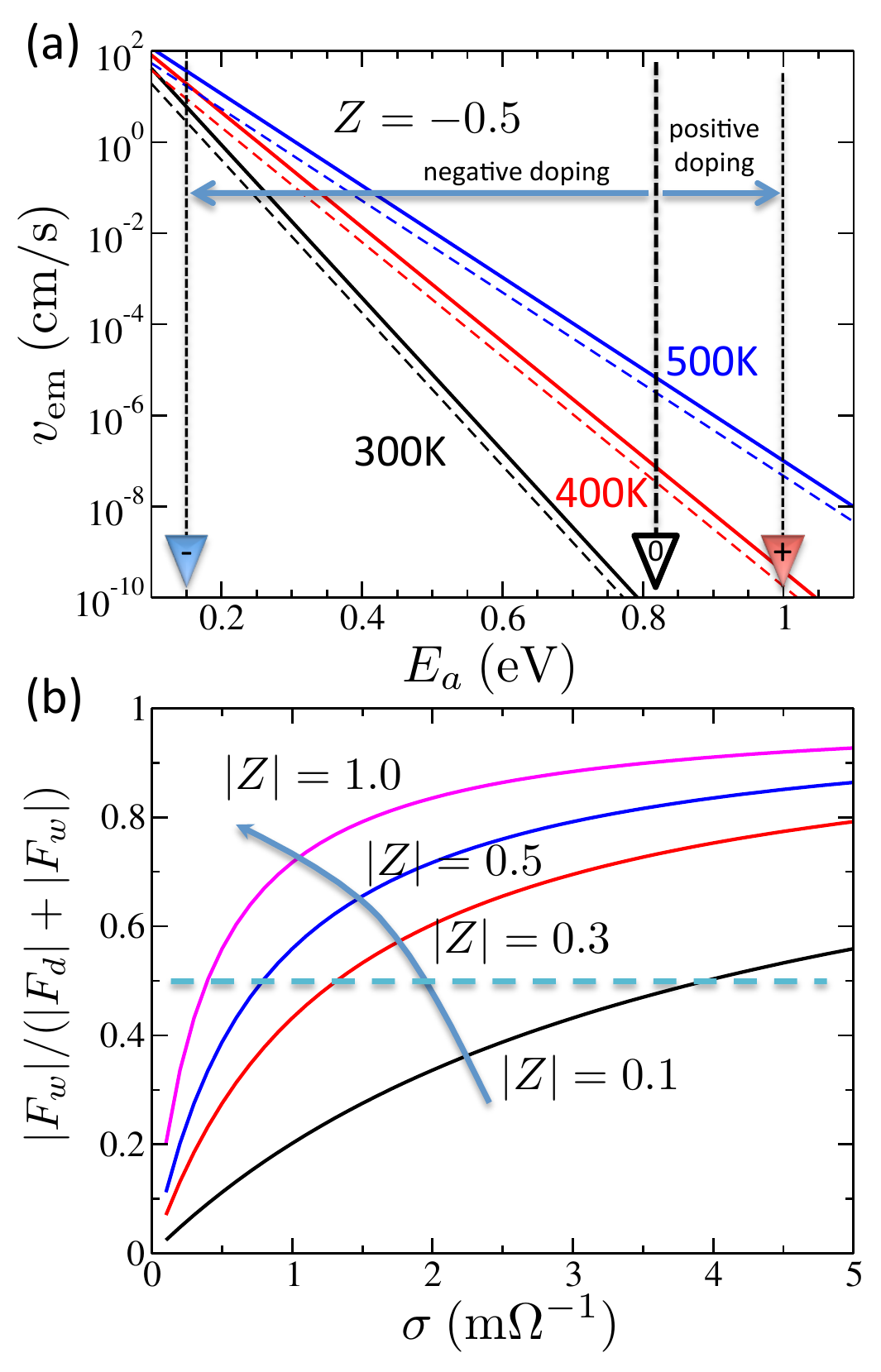}
\caption{\label{fig:vem} (a) The velocity of electromigration for $Z=-0.5$ versus the activation energy of hopping diffusion. The velocities for attempt frequencies of $\nu_0=30~{\rm THz}$ and $\nu_0=14~{\rm THz}$ are depicted by solid and dashed lines, respectively. (b) Relative contribution of the wind force to the total electromigration force (versus graphene conductivity) for a range of adsorbate charges.}
\end{figure}

Fig.~\ref{fig:vem}(b) shows the relative contribution of the wind force to the driving force
of the electromigration versus the conductivity of graphene.
The horizontal dashed line marks where the wind force and direct contributions become equal in magnitude. As is
seen, for an adsorbate charge of $Z=-0.5$ the wind force dominates
over the direct one for not very small conductivities. Interestingly, the two contributions to the driving force of electromigration have opposite signs at $Z>0$. At this condition, the dashed line marks where the driving force vanishes exactly since the direct and wind forces becomes equal in magnitude but opposite in sign. For some other type of adsorbates, e.g., alkali
metals, the magnitude of the charge localized on the adsorbate could
reach up to $|Z|\sim1$, so the wind force can become very strong. 

The velocity of electromigration can be found from the driving
force using the Einstein-Smoluchowski relation\cite{Kubo1985} as 
\begin{equation}
{\bf v}_{{\rm em}}=\frac{D}{k_{B}T}{\bf F}.
\end{equation}
Fig.~\ref{fig:vem}(a) shows the dependence of this velocity of oxygen
migration on graphene versus the activation energy of hopping diffusion
at the temperature ranging from $T=300$ K to $T=500$ K. The assumed parameters are $Z=-0.5$, $j=1$ A/mm and the conductivity
of graphene sample is taken $\sigma=1$ $m\Omega^{-1}$.  As is seen, the migration velocity can be quite high.

\section{Conclusion}\label{sec:concl}

We have presented detailed derivations of electromigration force acting on bivalent adsorbates on a surface of graphene. For such adsorbates we have demonstrated that the electron wind force can be accurately calculated using  a second-order perturbation theory with respect to the scattering potential (i.e., Born approximation). The scattering potential can be assumed to be just a Coulomb potential of a single point-wise charge within a graphene plane, the magnitude of the charge obtained as the partial charge of the adsorbate group together with the carbon atoms of graphene it is attached to. We have further shown that essentially all the lattice effects can be disregarded due to a smoothness of the scattering Coulomb potential on the size of graphene unit cell. Therefore, Dirac electrons in graphene can be treated as free particles being scattering by a Coulomb potential of the adsorbate. The only place where lattice effects enter the calculation of the electron wind force is the renormalization of the the scattering potential by the Bloch function overlap coefficients, Eq.~(\ref{eq:windf}). The physical reasoning behind this renormalization is that electron states in graphene are chiral and a smooth scattering potential (e.g., Coulomb) does not affect their pseudo spin resulting in the suppressed backscattering.\cite{CastroNeto2009-109} This renormalization, however, only introduces a factor on the order of $\sim 1$.

Using the developed formalism, we have evaluated an electromigration force and the electromigration drift velocity for two important examples of bivalent adsorbates: epoxy group ($\rm -O-$) and amino group ($\rm -NH-$). For these two functional groups we have used the accurate electronic structure theory calculations to assess their partial charge, as well as the attempt frequency and the energy barrier of diffusion. Using these parameters, the resulting velocities of electromigration were found to reach as high as $1-10~{\rm cm/s}$ for the epoxy group on a negatively doped graphene at room temperature, see Fig.~\ref{fig:vem}(a). Therefore, an electromigration of bivalent functional groups on graphene is expected to be efficient and observable at typical experimental conditions.

The biggest uncertainty in the present analysis of the electron wind force (as well as the direct force) is the value of the localized charge associated with an adsorbate. As it was already discussed above, this uncertainty is not related to any inaccuracies in electronic structure calculations but instead stems from the fundamental problem - impossibility to rigorously separate the charge density into current-carrying {\em mobile} and adsorbate-trapped {\em localized} components.\cite{Sorbello1997-159} A more rigorous although less physically transparent and more computationally intensive approach to evaluate the driving force of electromigration would be to directly calculate the total force acting on an adsorbate due to scattering of {\em all} the electrons within a current-carrying graphene.  

We thank Alejandro M. Suarez for multiple discussions and Josiah Bjorgaard for help with the manuscript. This work was performed under the NNSA of the U.S. DOE at LANL under Contract No. DE-AC52-06NA25396, and, in part,
by NRC/ONR.

\appendix

\section{Wind force from point defect\label{sec:point_def}}

The tight-binding Hamiltonian for pristine graphene (i.e., without
adsorbates) is given by Eq.~(\ref{eq:tb_0-1}). A covalently bound
adsorbate modifies the electronic structure of graphene by changing
the hybridization of certain number of carbon atoms in graphene. In
particular, when an atomic oxygen is attached to graphene in the equilibrium
epoxy configuration, it changes the hybridization of two carbon atoms
it is attached to from $sp^{2}$ to $sp^{3}$. The $sp^{3}$ hybridization
of a certain carbon atom due to the presence of an adsorbate can then
be approximately described by removing the corresponding $p_{z}$ orbital
from the tight-binding Hamiltonian. This is accomplished by adding
a scattering term $e\hat{U}_{\gamma}$ to $\hat{H}_{0}$ in Eq.~(\ref{eq:tb_0-1}),
where the scattering potential reads as
\begin{equation}
\hat{U}_{\gamma}=\sum_{k,l}U_{\gamma,kl}\hat{c}_{k}^{\dagger}\hat{c}_{l}\equiv (1-\gamma)\frac{t}{e}\sum_{(kl)\in1}\hat{c}_{k}^{\dagger}\hat{c_{l}}+\gamma\frac{t}{e}\sum_{(ij)\in2}\hat{c}_{k}^{\dagger}\hat{c}_{l}.\label{eq:U_point_def}
\end{equation}
The last two summations are performed with respect to hopping integrals
``turned off'' by the adsorbate in its initial, $(kl)\in1$, and
the final, $(kl)\in2$, position (see Fig.~\ref{fig:Schematic}). The role of the
position of the defect is played by $\gamma$, which, when changed
from $0$ to $1$, effectively ``shifts'' the defect from its initial
to its final position. 

The linear response treatment of the wind force with the defect potential
given by Eq.~(\ref{eq:U_point_def}) is done via the straightforward
generalization of Eq.~(\ref{eq:force_11_asymm}) by substituting
matrix elements of the adsorbate potential defined in real space with
those obtained for Eq.~(\ref{eq:U_point_def}). Specifically, the
wind force becomes
\begin{equation}
{\bf F}_{w}={\bf F}_{1,neq}^{-}=-\sum_{kl,mn}\left(\nabla_{{\bf R}}U_{\gamma,kl}\right)\delta\chi_{e}^{-}(kl;mn)U_{\gamma,mn}.\label{eq:wind_point}
\end{equation}
Here, $\nabla_{{\bf R}}$ has to be understood as $\left(\nabla_{{\bf R}}\gamma\right)\partial_{\gamma}$
where $\nabla_{{\bf R}}\gamma$ is approximately given by $\sim1/a$
in the case of an adsorbate hopping on distances on the order of the carbon-carbon
distance in graphene, which is of course the case considered in this
work.

The non-equilibrium electronic contribution to the linear response
function is given by \cite{Giuliani2005}
\begin{equation}
\delta\chi_{e}(kl;mn)=\frac{e^2}{\hbar}\sum_{ij}\frac{\delta f_{i}-\delta f_{j}}{\omega_{ij}+i\delta}\langle i|\hat{\rho}_{kl}|j\rangle\langle j|\hat{\rho}_{mn}|i\rangle,\label{eq:chi_point}
\end{equation}
where $\hat{\rho}_{kl}=\hat{c}_{k}^{\dagger}\hat{c}_{l}$ is the electron
density operator in the tight-binding representation. Eigenfunctions
of Eq.~(\ref{eq:tb_0-1}), i.e., Bloch waves of pristine graphene
without adsorbate, are denoted by $|i\rangle$ and $|j\rangle$. Eq.~(\ref{eq:chi_point})
obviously reduces to Eq.~(\ref{eq:chi_xx}) upon substitution $\hat{\rho}_{kl}\rightarrow\hat{\rho}({\bf x})$
and $\hat{\rho}_{mn}\rightarrow\hat{\rho}({\bf x}')$, where $\hat{\rho}({\bf x})=\hat{\varphi}^{\dagger}({\bf x})\hat{\varphi}({\bf x})$
is the real-space electron density operator. Accordingly, the antisymmetric
part of this response function is defined as $\delta\chi_{e}^{-}(kl;mn)=\left[\delta\chi_{e}(kl;mn)-\delta\chi(mn;kl)\right]/2$,
which yields
\begin{align}
\delta\chi_{e}^{-}(kl;mn)&=-i\frac{\pi e^2}{\hbar}\sum_{ij}\left[\delta f_{i}-\delta f_{j}\right]\delta(\omega_{ij})\nonumber\\
&\times \langle i|\hat{\rho}_{kl}|j\rangle\langle j|\hat{\rho}_{mn}|i\rangle.
\end{align}

Normalized eigenfunctions of $\hat{H}_{0}$ have local spatial amplitude
inversely proportional to the square root of the area of the graphene
sheet, and, therefore, the magnitude of an overlap factor, $\langle i|\hat{\rho}_{kl}|j\rangle$,
can be estimated as (in the rest of the appendix sign $\sim$ stands
for the approximate equality in \emph{magnitude})
\begin{equation}
\langle i|\hat{\rho}_{kl}|j\rangle\sim a^{2}/S,
\end{equation}
and its phase depends on the specific orientation of ${\bf k}_{i}$
and ${\bf k}_{j}$ and the orientation of the $(kl)$ link. In particular,
at large distances between pairs $(mn$) and $(kl)$ the phase of
the overlap factor can oscillate rapidly as a function of  ${\bf k}_{i}$
and ${\bf k}_{j}$ resulting in very small $\chi(kl;mn)$. At short distances,
however, the overlap factors simply produce a prefactor on the order
of one in front of the Lindhard function, so that $\delta\chi_{e}^{-}(mn;kl)\sim\frac{a^{4}e^2}{\hbar S^{2}}\sum_{ij}\left[\delta f_{i}-\delta f_{j}\right]\delta(\omega_{ij})$
and the wind force becomes
\begin{equation}
{\bf F}_{w}\sim\frac{a^{4}e^2}{\hbar S^{2}}\sum_{kl,mn}\left(\nabla_{{\bf R}}U_{\gamma,kl}\right)\sum_{ij}\left[\delta f_{i}-\delta f_{j}\right]U_{\gamma,mn}\delta(\omega_{ij}).\label{eq:wind_point2}
\end{equation}
Using the approximate identity $\nabla_{{\bf R}}U_{\gamma,kl}\sim a^{-1}\partial_{\gamma}U_{\gamma,kl}$
(see above) and that $\partial_{\gamma}U_{\gamma,kl}\sim U_{\gamma,kl}$,
as follows from Eq.~(\ref{eq:U_point_def}), we rewrite the expression
for the wind force as
\begin{equation}
{\bf F}_{w}\sim\frac{a^{4}e^2}{\hbar S^{2}}\sum_{ij}\sum_{kl,mn}\left[\delta f_{i}-\delta f_{j}\right]\frac{1}{a}U_{\gamma,kl}U_{\gamma,mn}\delta(\omega_{ij}),\label{eq:wind_point3}
\end{equation}
where $1/a$ within the summation plays the role of momentum transfer
factor, ${\bf k}_{i}-{\bf k}_{j}-{\bf G}$, in Eq.~(\ref{eq:Fw_general}),
since it originates from the same differentiation of the adsorbate
potential with respect to its position ${\bf R}$. To proceed further,
we recall that (i) $U_{\gamma,kl}\sim U_{\gamma,mn}\sim t/e$ [see Eq.~(\ref{eq:U_point_def})], and that
(ii) the summation with respect to $k$,$l$, $m$ and $n$ runs over
a finite small subset of indices corresponding to turned-off hopping
integrals for a particular adsorbate. The latter statement implies
that $\sum_{kl,mn}$ produces a factor on the order of one. We thus
can rewrite Eq.~(\ref{eq:wind_point3}) as
\begin{equation}
{\bf F}_{w}\sim\frac{e^2}{\hbar S^{2}}\sum_{ij}\left[\delta f_{i}-\delta f_{j}\right]\frac{1}{a}\left(a^{2}t/e\right)^{2}\delta(\omega_{ij}).
\end{equation}
Comparing this expression with Eq.~(\ref{eq:wind_free}) one concludes
that the scattering by such a defect potential of the adsorbate can
be thought of as a scattering of jellium electrons (i.e., local field
effects are neglected) in graphene by a delta-function potential with
an amplitude of $\sim a^{2}t/e$, i.e., one can approximately substitute
$\hat{U}_{\gamma}$ in Eq.~(\ref{eq:U_point_def}) with $U_{{\bf R}}({\bf x})\approx\alpha \frac{a^{2}t}{e}\delta({\bf x}-{\bf R})$,
where $\alpha$ is the prefactor on the order of one.

\section{Strong adsorbate scattering\label{app:strong}}

To assess the contribution of higher order scattering events, i.e., those beyond what is already included in ${\bf F}_{1,neq}^{-}$ in
Eq.~(\ref{eq:tforce}), we estimate the renormalization (or dressing)
of the adsorbate single-particle potential by multiple scattering events.  Resummation of the adsorbate-induced electron scattering events to all orders is conveniently done using a general formalism of single-particle Green's functions. Symbolically, an exact electron Green's function, $G$, for a system with an impurity can be written as \cite{Economou1990}
\begin{equation}
G=g+gH_1g+gH_1gH_1g+...,\label{eq:Dyson_series}
\end{equation}
where $g$ is the {\em bare} Green's function for the system without the impurity and $H_1$ is the perturbation operator due to the presence of the impurity. One can introduce a {\em dressed} perturbation operator as $\tilde{H}_1=H_1+H_1gH_1+H_1gH_1gH_1+...$ , so that the exact Green's function reads as
\begin{equation}
G=g+g\tilde{H}_1 g.\label{eq:dressedH1}
\end{equation}
This expression looks similar to the first-order Born approximation\cite{Economou1990} except that the perturbation operator in the former is {\em dressed} and, therefore, Eq.~(\ref{eq:dressedH1}) is exact. Reformulation of the scattering theory in terms of the dressed perturbation operator, also known as $T$-matrix,\cite{Economou1990} is very convenient since once this operator is found, the first-order Born approximation machinery can be used to get an {\em exact} result accounting for scattering events to all orders. Our goal in this section is to find the dressed scattering potential for adsorbates considered in this work. Once it is found, it can be directly used in Eq.~(\ref{eq:rho_linear}).

In this work, the deviation from equilibrium is considered only up to linear terms in the current, and the current
already enters Eq.~(\ref{eq:windf}). Therefore, the renormalization of the adsorbate potential can be considered at
equilibrium, so we use the standard zero-temperature Green's function
formalism.\cite{Economou1990,Mahan2000} Since the inter-cone scattering due to
the Coulomb potential of an adsorbate is suppressed (see Sec.~\ref{sub:effective_potential}),
we consider only scattering within a single Dirac cone, ${\bf K}$
for definiteness. The Lippmann-Schwinger
equation for the exact Green's function reads as
\begin{equation}
\hat{G}(\omega)=\frac{\hbar^{-1}}{\omega-\hat{H}+i\delta}=\hat{g}(\omega)+\hat{g}(\omega)\hat{H}_{1}\hat{G}(\omega),
\end{equation}
where $\hat{H}_{1}$ denotes the operator of the Coulomb potential
of the adsorbate. The exact Green's function, $\hat{G}(\omega)$,
and the bare one, $\hat{g}(\omega$), the latter specifying the response
of the pristine graphene without adsorbates, are defined in the full Hilbert space of
the problem. We now project this expression onto the normalized plane
wave basis, $|{\bf k}\rangle$, defined through $\langle{\bf x}|{\bf k}\rangle=S^{-1/2}e^{i{\bf k}{\bf x}}$. The resulting Lippmann-Schwinger equation is
\begin{equation}
\hat{G}_{{\bf k}{\bf k'}}(\omega)=\hat{g}_{\bf k}(\omega)\delta_{{\bf k}{\bf k}'}
+\sum_{\bf q} \hat{g}_{\bf k}(\omega)\hat{H}_{1,{\bf k}{\bf q}}\hat{G}_{{\bf q}{\bf k}'}(\omega), \label{eq:LSproj}
\end{equation}
where all the Green's functions at fixed quasimomentum indices are $2\times2$ matrices, consistently with the spinor representation of the Bloch functions in graphene, Eq.~(\ref{eq:Bloch}). The bare projected Green's function is given by
\begin{align}
\hat{g}_{{\bf k}}(\omega)&=\frac{\hbar^{-1}}{\omega-\hat{H}_{{\bf K}}(k)+i\delta}\nonumber\\
&=\sum_{b=\pm1}|u_{b,{\bf K}}({\bf k})\rangle\frac{\hbar^{-1}}{\omega-bv_F k+i\delta}\langle u_{b,{\bf K}}({\bf k})|,
\end{align}
where $b=+1$ ($-1$) for an electron state in $\pi^{*}$ ($\pi$)
band and the Bloch functions in the spinor representation are given
in Eq.~(\ref{eq:Bloch}).

The operator of the Coulomb interaction in the contact form is given
by
\begin{equation}
\hat{H}_{1,{\bf k}{\bf k}'}=\frac{e}{S}\hat{U}_{C,{\bf k}{\bf k}'}^{c}=\frac{e}{S}U_{C}^{c}\hat{\sigma}_{0},\label{eq:H1}
\end{equation}
where $\hat{\sigma}_0$ is the $2\times 2$ unit matrix and $U_{C}^{c}$ is the amplitude of the Coulomb potential in the
contact form, given by Eq.~(\ref{eq:Ccontact}). Expanding the Lippmann-Schwinger equation (\ref{eq:LSproj}) into the Dyson series and using Eq.~(\ref{eq:H1}) one obtains
\begin{gather}
\hat{G}_{{\bf k}{\bf k}'}(\omega)=\hat{g}_{\bf k}(\omega)\delta_{{\bf k}{\bf k}'}
+\frac{e}{S}\hat{g}_{\bf k}(\omega)\hat{U}^c_{C,{\bf k}{\bf k}'}\hat{g}_{{\bf k}'}(\omega)\nonumber\\
+\frac{e^2}{S^2}\sum_{\bf q}\hat{g}_{\bf k}(\omega)\hat{U}^c_{C,{\bf k}{\bf q}}\hat{g}_{{\bf q}}(\omega)
\hat{U}^c_{C,{\bf q}{\bf k}'}\hat{g}_{{\bf k}'}(\omega)+...
\label{eq:Dexp}
\end{gather}
Defining
\begin{gather}
\hat{\mathcal{U}}_{{\bf k}{\bf k}'}(\omega)=\hat{U}_{C,{\bf k}{\bf k}'}^{c}+\frac{e}{S}\sum_{{\bf q}}\hat{U}_{C,{\bf k}{\bf q}}^{c}\hat{g}_{{\bf q}}(\omega)\hat{U}_{C,{\bf q}{\bf k}'}^{c}\nonumber\\
+\frac{e^2}{S^{2}}\sum_{{\bf q},{\bf q}'}\hat{U}_{C,{\bf k}{\bf q}}^{c}\hat{g}_{{\bf q}}(\omega)\hat{U}_{C,{\bf q}{\bf q}'}^{c}\hat{g}_{{\bf q}'}(\omega)\hat{U}_{C,{\bf q}'{\bf k}'}^{c}+...
\label{eq:Dexp1}
\end{gather}
one readily sees that Eq.~(\ref{eq:Dexp}) becomes
\begin{equation}
\hat{G}_{{\bf k}{\bf k}'}(\omega)=\hat{g}_{\bf k}(\omega)\delta_{{\bf k}{\bf k}'}
+\frac{e}{S}\hat{g}_{\bf k}(\omega)\hat{\mathcal U}^c_{C,{\bf k}{\bf k}'}(\omega)\hat{g}_{{\bf k}'}(\omega)
\end{equation}
Therefore, $\hat{\mathcal U}^c_{C,{\bf k}{\bf k}'}(\omega)$ is the dressed scattering potential (or $T$-matrix).

An important observation here is that the bare Coulomb potential, ${\hat U}^c_{C,{\bf k}{\bf k}'}$,
is proportional to the unit matrix, $\hat{\sigma}_{0}$. The bare Green's
function, $\hat{g}_{{\bf k}}(\omega)$, becomes
proportional to $\hat{\sigma}_{0}$ once averaged over the angular part of ${\bf q}$, as follows from Eq.~(\ref{eq:uu_aver}). Furthermore, $\hat{U}^c_{C,{\bf k}{\bf k}'}$ does not actually depend on quasimomenta, as follows from Eq.~(\ref{eq:H1}). The only dependence on quasimomenta in the r.h.s. of Eq.~(\ref{eq:Dexp1}) (i.e., via $\hat{g}_{\bf q}(\omega)$) is ``integrated out". These two observations make it possible to write down the dressed potential as $\hat{\mathcal{U}}_{{\bf k}{\bf k}'}(\omega)=\mathcal{U}(\omega)\hat{\sigma}_{0}$,
where
\begin{align}
\mathcal{U}(\omega)&=U^c_C+U^c_Cg(\omega)U^c_C+U^c_Cg(\omega)U^c_Cg(\omega)U^c_C+...\nonumber\\
&=\frac{U^c_C}{1-U^c_Cg(\omega)},
\end{align}
and 
\begin{equation}
g(\omega)=\frac{e}{4\pi\hbar}\int_{0}^{q_{*}}dq\, q\left[\frac{1}{\omega-v_F q+i\delta}+\frac{1}{\omega+v_F q+i\delta}\right],
\end{equation}
The two terms in the integrand correspond to conduction and valence
bands. The real part of the integral is calculated as
\begin{align}
g'(\omega)&=\frac{e}{2\pi\hbar}P\int_{0}^{q_{*}}dq\,\frac{\omega q}{\omega^{2}-v_F^{2}q^{2}}\nonumber\\
&=-\frac{e\omega}{4\pi \hbar v_F^{2}}\ln\left|\frac{v_F^{2}q_{*}^{2}-\omega^{2}}{\omega^{2}}\right|.
\end{align}
The imaginary part is given by
\begin{equation}
g''(\omega)=-\frac{e\left|\omega\right|}{4\hbar v_F^{2}},
\end{equation}
where it is assumed that $|\omega|\ll\omega_{*}\equiv v_F q_{*}$ - the
cutoff frequency. For the Coulomb potential it is reasonable to associate
the cutoff frequency with the Debye screening length through $\omega_{*}=v_F q_{D}$.
As a result, the amplitude of the dressed scattering potential becomes
\begin{equation}
\mathcal{U}(\omega)=\frac{U^c_C}{1+\frac{e\omega U^c_C}{4\pi \hbar v_F^{2}}\ln\left|\frac{v_F^{2}q_{D}^{2}-\omega^{2}}{\omega^{2}}\right|+i\frac{e|\omega| U^c_C}{4\hbar v_F^{2}}}.\label{eq:Udressed}
\end{equation}
To assess the extent of the renormalization, we take $\omega=v_F q_{F}$
and compare the real terms in the denominator of this expression. It
is now seen that before, in Eq.~(\ref{eq:UC_unitless}), we have correctly
identified all the dimensional factors, but not the
dimensionless one, $\frac{1}{4\pi}\ln\left|\frac{q_{D}^{2}-q_{F}^{2}}{q_{F}^{2}}\right|$,
which depends slightly on the nature of the dielectric environment
via $q_{D}$. For the ${\rm SiO_{2}}$ substrate in vacuum, this factor
becomes $\approx1/2\pi$, so that
\begin{equation}
\lambda/2\pi\approx0.02-0.08.
\end{equation}
This, unlike simply $\lambda$, can be safely considered to be much smaller than $1$, validating our assumption that for adsorbates considered
in this work the first non-vanishing contribution to the wind force with
the respect to the scattering potential, ${\bf F}_{1,neq}^{-}$ is
enough to accurately evaluate the electron wind force.

\section{Diffusion coefficient for hopping of bivalent adsorbates on graphene\label{sec:diff}}

Diffusion on two-dimensional (2D) surface without sources is governed by 
\begin{gather}\label{eq:app:diff:contin-eq}
\partial_t \rho({\bf r},t) 
=
 - \nabla\cdot {\bf j}({\bf r},t),
\\\label{eq:app:diff:flux}
j_i 
=
 - \sum_j D_{ij}({\bf r}) \partial_{r_j} \rho({\bf r},t).
\end{gather}
When the system is isotropic the tensor of diffusion coefficients reduces to a single number, i.e. $D_{ij}=D\delta_{ij}$. When diffusion is anisotropic the diffusion tensor has two distinct eigenvalues ${\frak b}_1\neq {\frak b}_2$. In orthonormal (Cartesian) frame of reference the diagonalization, $\hat D {\frak v}_i = {\frak b}_i {\frak v}_i$, performs rotation of the frame of reference to the eigenbasis ${\frak v}_i\cdot {\frak v}_j = \delta_{ij}$ of $\hat D$. It is straightforward to see that when $D_{ij}$ does not depend on ${\bf r}$, anisotropic diffusion (${\frak b}_1\neq {\frak b}_2$) takes place when the rotation symmetry of the system is not greater than $C_2$. Systems with higher rotational symmetry will cause isotropic diffusion. 

For diffusion of a single particle on a macroscopically uniform surface ($D_{ij}$ are constants) it is appropriate to chose
\begin{eqnarray}\label{eq:app:diff:ic}
\rho({\bf r},0)=\delta({\bf r}-{\bf r}_0),
\end{eqnarray}
as the initial condition. In this case Eqs.~(\ref{eq:app:diff:contin-eq}) and (\ref{eq:app:diff:flux}) can be solved exactly
\begin{eqnarray}\label{eq:app:diff:solution}
\rho({\bf r},t) = \frac{1}{4\pi t\sqrt{{\frak b}_1{\frak b}_2}} \exp\left[
- \sum_{i=1,2}\frac{(\Delta{\bf r}\cdot {\frak v}_i)^2}{4{\frak b}_i t} 
\right],
\end{eqnarray}
where $\Delta {\bf r} = {\bf r}-{\bf r}_0$. Calculating the second moments of of this distribution we find 
\begin{eqnarray}\label{eq:app:diff:moments}
D_{ij} = \frac{1}{2t}\langle \Delta r_i \Delta r_j \rangle.
\end{eqnarray}

For simple lattices with one site per unit cell (Bravais lattices) the diffusion coefficient can be obtained by simply mapping Eqs.~(\ref{eq:app:diff:contin-eq}) and (\ref{eq:app:diff:flux}) on to the lattice
\begin{eqnarray}\label{eq:app:diff:Pi}
\partial_t P_i = \sum_{\langle ij\rangle} 
[
\Gamma_{ji} P_j - \Gamma_{ij} P_i
],
\end{eqnarray}
where $P_i$ is the probability to occupy site $i$, $\langle ij\rangle$ denotes summation over the nearest neighbors, and $\Gamma_{ij}$ are the rates for the particle to jump to one of its neighbors (probability of jump per unit time). The correspondence between the continuous limit of Eqs.~(\ref{eq:app:diff:Pi}) and (\ref{eq:app:diff:contin-eq}-\ref{eq:app:diff:flux}) gives the values of $\hat D$.

\begin{figure}
\includegraphics[width=0.8\columnwidth]{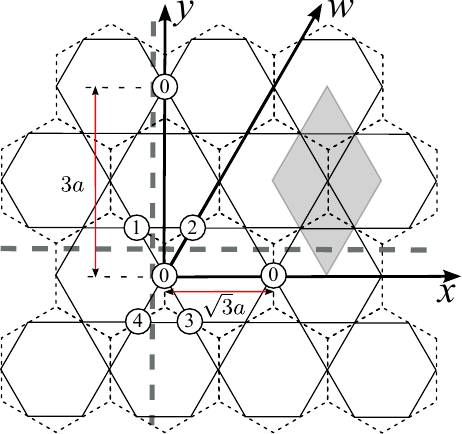}
\caption{\label{fig:o-pos}
The lattice of adsorption sites for bivalent adsorbates on graphene. Dashed lines represent the underlying graphene lattice. Solid lines show the lattice of adsorption sites -- symmetric Kagome lattice. The unit cell of the Kagome lattice is highlighted by the shaded area. It contains three non-equivalent sites, such as (0), (1), and (2).
}
\end{figure}

In the case of non-Bravais lattices, the unit cell contains several sites, and an adsorbate can perform random jumps within the unit cell as well. As a result, Eq.~(\ref{eq:app:diff:Pi}) has to be formulated for each non-equivalent lattice point, which produces a set of coupled differential (finite elements) equations that cannot be easily transformed into Eqs.~(\ref{eq:app:diff:contin-eq}-\ref{eq:app:diff:flux}) in the macroscopic limit.\cite{Braun1998-14870} Adsorption sites for bivalent adsorbates considered in this work form symmetric Kagome lattice [see Fig.~\ref{fig:o-pos}]. This lattice has three non-equivalent lattice sites in the unit cell and, thus, Eq.~(\ref{eq:app:diff:Pi}) cannot be used directly.

The macroscopic diffusion equation can be most easily recovered by performing a random walk through the lattice of adsorption sites numerically evaluating the right-hand side of Eq.~(\ref{eq:app:diff:moments}). We perform random walk (Monte Carlo) calculations that models the system of a single bivalent adsorbate on graphene. At each time step adsorbate is allowed to jump with small probability ${\cal P}$. The probability to jump to any of the four neighbor sites is therefore $4{\cal P}$, while the probability to stay on the current lattice site is $1-4{\cal P}$. This corresponds to the choice of discrete time $t = t_0 N$, where $N$ is a number of steps in a random walk trajectory and $t_0$ is the time for a single step. This time is related to a hopping rate $\Gamma$ [Eq.~(\ref{eq:diff_rate})] via $\Gamma t_0 = {\cal P}$. The specific choice of ${\cal P}$ is not important.

\begin{figure}
\includegraphics[width=0.99\columnwidth]{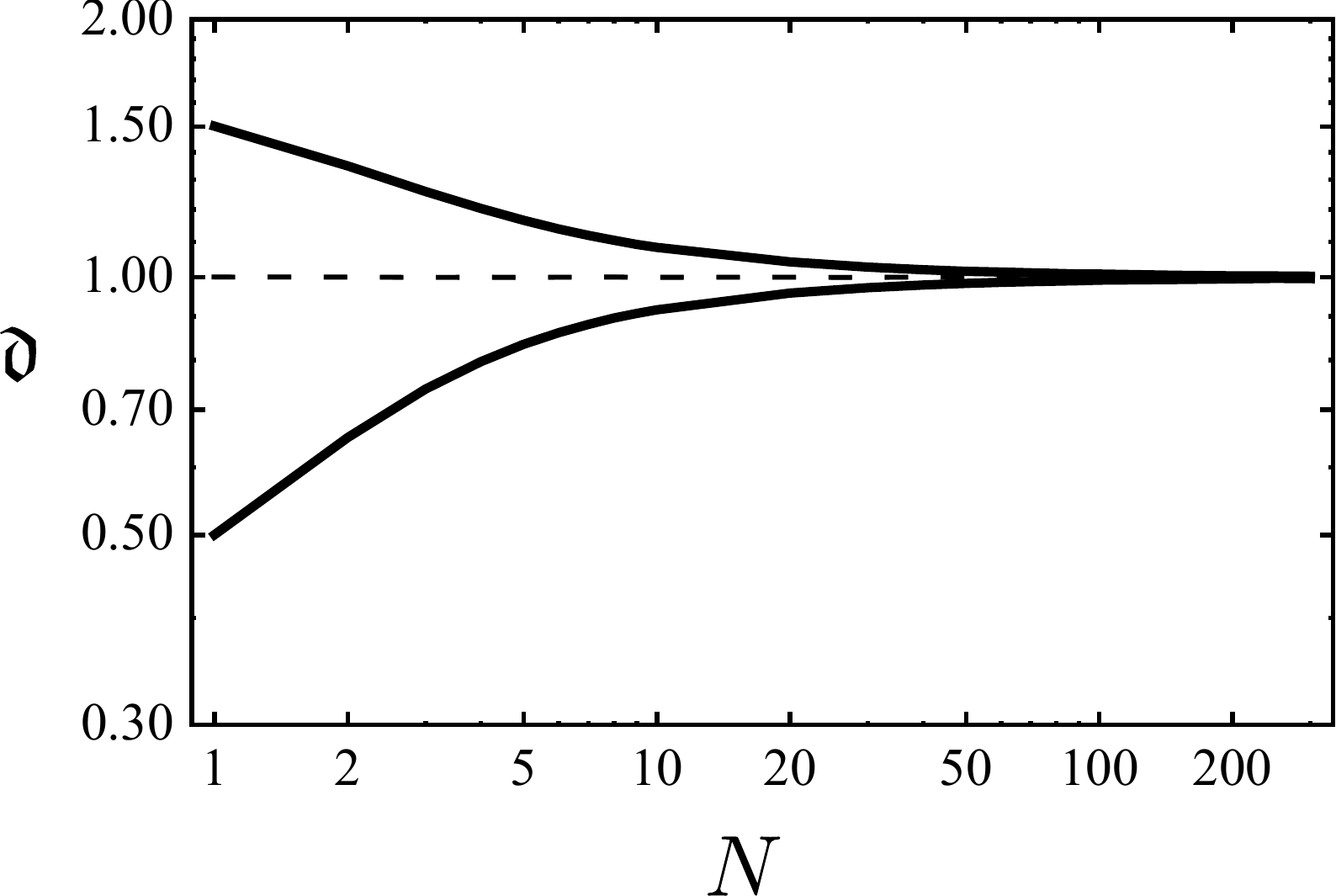}
\caption{\label{fig:diff}
Numerical results for random walk on Kagome lattice. Solid curves are the eigenvalues ${\frak d}$ of matrix ${\frak D}$ in Eq.~(\ref{eq:app:diff:M}) as a function of $N$. The curves are the same for different starting positions within the unit cell of the lattice. The diffusion distribution for each starting position has $C_2$ symmetry of the Kagome lattice site. The diffusion matrix for non-identical lattice sites are related by $2\pi/3$ rotations. The diffusion approaches the isotropic limit at large scales, i.e. $N\to\infty$. If ${\frak D}$ is averaged over the three non-equivalent starting positions, dashed curves, the symmetry is raised to $C_3$ and ${\frak D}$ becomes isotropic already at $N=1$. In both cases $10^7$ walks (trajectories) are taken to recover the probability distributions. 
}
\end{figure}

In Fig.~\ref{fig:diff} we plot the {\it eigenvalues}, ${\frak d}$, of the dimensionless microscopic diffusion tensor
\begin{eqnarray}\label{eq:app:diff:M}
{\frak D} = 
\frac{1}{d^2}
\frac{1}{2{\cal P}N}
\left(
\begin{array}{cccc}
\langle \Delta x^2 \rangle & \langle \Delta x \Delta y \rangle\\
\langle \Delta y \Delta x \rangle & \langle \Delta y^2 \rangle\\
\end{array}
\right),
\end{eqnarray}
where $d=\sqrt{3}a/2$ is the hopping distance -- the distance between nearest cites in the Kagome lattice. At small $N$ the walk is anisotropic. At larger $N$ both eigenvalues ${\frak d}_{i=1,2}\to 1$, indicating isotropic macroscopic diffusion,
\begin{eqnarray}\label{eq:app:diff:D}
D_{ij} = d^2\Gamma{\frak D}_{ij} = d^2\Gamma\delta_{ij} = \frac{3a^2}{4} \Gamma\delta_{ij}.
\end{eqnarray}

Note that there are three non-equivalent initial positions for the walker withing the unit cell. Each such position has rotational symmetry $C_2$, and, hence, ${\frak D}$ is anisotropic for finite $N$ (${\frak d}_1\neq {\frak d}_2$, see solid curves in Fig.~\ref{fig:diff}) However, when $N\to\infty$, we obtain an isotropic diffusion tensor, i.e. ${\frak d}_1\to {\frak d}_2$. Indeed, addition of one more jump (or even several jumps) to the walking trajectory cannot change the final probability distribution. Therefore at $N\to\infty$ all three points within the unit cell become equivalent and the symmetry is raised to $C_3$ resulting in an isotropic diffusion tensor. The same isotropic result can be obtained for small $N$ (starting with $N=1$ for this specific lattice) if we average random walk with respect to the starting position (see the dashed line in Fig.~\ref{fig:diff}). Due to this dramatic simplification, it is possible to obtain correct macroscopic diffusion coefficient for the symmetric Kagome lattice analytically by performing one step ($N=1$) random walk, while performing the average over the staring positions of the walker. There are several ways to perform such averaging.

\subsection{Central limit theorem}
\label{subsec:CLT}

The first and the most straightforward way is to write the probability distribution for the four nearest neighbors after one step of the random walk. Since the probability for the adsorbate to jump to any of the four neighbors is the same, if we start from position 0, shown in Fig.~\ref{fig:o-pos}, the probability distribution at sites 1 through 3 becomes
\begin{equation}
P_4({\bf r})
\sim 
\frac{1}{4}\sum_{i=1}^4\delta({\bf r}-{\bf r}_{i}),
\end{equation}
where ${\bf r}_i=\pm\frac{\sqrt{3}a}{4}{\bf e}_x\pm\frac{3a}{4}{\bf e}_y$. Based on this distribution we obtain
\begin{eqnarray}\label{eq:app:diff:AV}
\langle \Delta r_i\Delta r_j \rangle_4
=
\left(
\begin{array}{cccc}
3a^2/16&0\\
0&9a^2/16\\
\end{array}
\right).
\end{eqnarray}
where $i$ ($j$) are $x$ or $y$. Note that $P_4(0)$, which defines the dependence on ${\cal P}$, is irrelevant because $\Delta{\bf r}=0$ at that point. If we repeat this calculations for the other two sites of the unit cell, positions 1 and 2, we obtain the same $\langle \Delta r_i\Delta r_j \rangle_4$ rotated by $2\pi/3$ or $-2\pi/3$ respectively. Averaging $\langle \Delta r_i\Delta r_j \rangle_4$ with respect to these three non-equivalent initial positions we have
\begin{eqnarray}\label{eq:app:diff:AVAV}
\langle\langle \Delta r_i\Delta r_j \rangle_4\rangle_3
=
\left(
\begin{array}{cccc}
3a^2/8&0\\
0&3a^2/8\\
\end{array}
\right).
\end{eqnarray}
Note that in this derivation the single (time) step involves jumps to all four neighbors, therefore $t= 1/4\Gamma$. Substituting Eq.~(\ref{eq:app:diff:AVAV}) into Eq.~(\ref{eq:app:diff:moments}) we obtain $D = d^2\Gamma$ as in Eq.~(\ref{eq:app:diff:D}).

The same result is obtained for an infinite walk by virtue of the central limit theorem. After $M$ steps described above the distribution of total hopping distances is $\sim e^{-4r^2/3a^2M}$. Note, once again, that at each such step all four neighbors are involved, thus $M=4\Gamma t$. Comparing this distribution with (\ref{eq:app:diff:solution}) we obtain $D = 3a^2\Gamma/4$ as in Eq.~(\ref{eq:app:diff:D}).

We emphasize that averaging over the three non-equivalent cites of the Kagome lattice is critical. Such averaging might not be valid for more complex non-Bravais lattices that are less connected or have larger number of sites per unit cell. Note also that because $\Gamma$ specifies a rate of leaving a cite via a {\em specific} link (see Subsection \ref{subsec:Hopping}), the total rate of leaving a cite of the Kagome lattice is given by $\Gamma_{tot}=4\Gamma$. This results in equality $M=4\Gamma t$ above. Using $\Gamma_{tot}=\Gamma$ underestimates the diffusion coefficient by a factor of four.

\subsection{Uniform current along $y$}

The same averaging over the non-equivalent lattice sites can be performed implicitly by calculating a uniform current through the system due to, e.g., some external force. We write the flow of probability in $y$ direction assuming translation symmetry in $x$ direction. This amounts to the flow through the thick dashed horizontal line in Fig.~\ref{fig:o-pos}. The current is given by $I_y = \Gamma (P_0-P_1) + \Gamma (P_0-P_2)$ per $x$-translation [see Fig.~\ref{fig:o-pos}]. The corresponding current density is $j_y = [2\Gamma (P_0-P_1)]/\sqrt{3}a$. The site population probability can be related to the continuous density as
$P_i=\rho({\bf r}_i)S_{K}$, where $S_{K}=\sqrt{3}a^{2}/2$
is the area per cite of the Kagome lattice. Therefore,
we obtain $j_y = \Gamma a [\rho({\bf r}_0)-\rho({\bf r}_0)]$. In the macroscopic limit the difference $\rho({\bf r}_0)-\rho({\bf r}_1)$ gives the gradient of probability density in the $y$ direction, i.e. $(3a/4)\partial_y \rho$. Therefore we obtain
\begin{equation}\label{eq:app:diff:jy}
j_y = -D_{yy}\partial_y\rho = -\Gamma\frac{3a^{2}}{4}\partial_y\rho.
\end{equation}

\subsection{Uniform current along $x$}

Similarly, we can consider the current in the $x$ direction that is uniform with respect to the $y$-translation. This amounts to the flow though the thick dashed vertical line in Fig.~\ref{fig:o-pos}). We obtain $I_x = [2\Gamma (P_0-P_2) + 2\Gamma (P_1-P_2)]/3a$ and $j_x = \Gamma a [\rho({\bf r}_0)-\rho({\bf r}_2) + \rho({\bf r}_1)-\rho({\bf r}_2)]/\sqrt{3}$. In the macroscopic limit $\rho({\bf r}_0)-\rho({\bf r}_2)\to(\sqrt{3}a/4)\partial_y \rho$ and $\rho({\bf r}_1)-\rho({\bf r}_2)\to(\sqrt{3}a/2)\partial_y \rho$, and we obtain
\begin{equation}\label{eq:app:diff:jx}
j_x = -D_{xx}\partial_x\rho =  -\Gamma\frac{3a^{2}}{4}\partial_x\rho.
\end{equation}
Both Eq.~(\ref{eq:app:diff:jx}) and (\ref{eq:app:diff:jy}) suggest the expression for the diffusion coefficient given in Eq.~(\ref{eq:app:diff:D}). 

\subsection{Uniform current along $w$}

The same coefficient also follows from Eq.~(\ref{eq:app:diff:Pi}) written for a single crystallographic direction, $w$ (see Fig.~\ref{fig:o-pos}), if we assume no current of probability in the perpendicular direction,
\begin{eqnarray}\label{eq:app:diff:PiW}
\partial_t P_0 = \Gamma P_2 + \Gamma P_4 - 2\Gamma P_0.
\end{eqnarray}
In the macroscopic limit the right-hand side approaches $-\partial_w j_w/S_K$, with
\begin{eqnarray}\label{eq:app:diff:jw}
j_w = -\frac{3a^2}{4}\Gamma\partial_w\rho({\bf r})
\end{eqnarray}
and the same value of $D$ is recovered. 

In all the above analytical examples the average over the three non-equivalent sites is taken explicitly or implicitly. In the case of symmetric Kagome lattice such averaging yields the correct macroscopic diffusion coefficient (\ref{eq:app:diff:D}). We should emphasize, however, that the above analytical derivations will not necessarily lead to the correct result for other lattices: a numerical random walk has to be performed. Note also that the expression for $D$ in Eq.~(\ref{eq:app:diff:D}) differs from the one used in Refs.~\onlinecite{Suarez2011-146802,Solenov2012-095504}, which might be due to the reason outlined at the very end of Subsection~\ref{subsec:CLT} above.


\begin{thebibliography}{65}%
\makeatletter
\providecommand \@ifxundefined [1]{%
 \@ifx{#1\undefined}
}%
\providecommand \@ifnum [1]{%
 \ifnum #1\expandafter \@firstoftwo
 \else \expandafter \@secondoftwo
 \fi
}%
\providecommand \@ifx [1]{%
 \ifx #1\expandafter \@firstoftwo
 \else \expandafter \@secondoftwo
 \fi
}%
\providecommand \natexlab [1]{#1}%
\providecommand \enquote  [1]{``#1''}%
\providecommand \bibnamefont  [1]{#1}%
\providecommand \bibfnamefont [1]{#1}%
\providecommand \citenamefont [1]{#1}%
\providecommand \href@noop [0]{\@secondoftwo}%
\providecommand \href [0]{\begingroup \@sanitize@url \@href}%
\providecommand \@href[1]{\@@startlink{#1}\@@href}%
\providecommand \@@href[1]{\endgroup#1\@@endlink}%
\providecommand \@sanitize@url [0]{\catcode `\\12\catcode `\$12\catcode
  `\&12\catcode `\#12\catcode `\^12\catcode `\_12\catcode `\%12\relax}%
\providecommand \@@startlink[1]{}%
\providecommand \@@endlink[0]{}%
\providecommand \url  [0]{\begingroup\@sanitize@url \@url }%
\providecommand \@url [1]{\endgroup\@href {#1}{\urlprefix }}%
\providecommand \urlprefix  [0]{URL }%
\providecommand \Eprint [0]{\href }%
\providecommand \doibase [0]{http://dx.doi.org/}%
\providecommand \selectlanguage [0]{\@gobble}%
\providecommand \bibinfo  [0]{\@secondoftwo}%
\providecommand \bibfield  [0]{\@secondoftwo}%
\providecommand \translation [1]{[#1]}%
\providecommand \BibitemOpen [0]{}%
\providecommand \bibitemStop [0]{}%
\providecommand \bibitemNoStop [0]{.\EOS\space}%
\providecommand \EOS [0]{\spacefactor3000\relax}%
\providecommand \BibitemShut  [1]{\csname bibitem#1\endcsname}%
\let\auto@bib@innerbib\@empty
\bibitem [{\citenamefont {Geim}\ and\ \citenamefont
  {Novoselov}(2007)}]{Geim2007-183}%
  \BibitemOpen
  \bibfield  {author} {\bibinfo {author} {\bibfnamefont {A.~K.}\ \bibnamefont
  {Geim}}\ and\ \bibinfo {author} {\bibfnamefont {K.~S.}\ \bibnamefont
  {Novoselov}},\ }\href@noop {} {\bibfield  {journal} {\bibinfo  {journal}
  {Nature Mat.}\ }\textbf {\bibinfo {volume} {6}},\ \bibinfo {pages} {183}
  (\bibinfo {year} {2007})}\BibitemShut {NoStop}%
\bibitem [{\citenamefont {{Lee}}\ \emph {et~al.}(2008)\citenamefont {{Lee}},
  \citenamefont {{Wei}}, \citenamefont {{Kysar}},\ and\ \citenamefont
  {{Hone}}}]{Lee2008-385}%
  \BibitemOpen
  \bibfield  {author} {\bibinfo {author} {\bibfnamefont {C.}~\bibnamefont
  {{Lee}}}, \bibinfo {author} {\bibfnamefont {X.}~\bibnamefont {{Wei}}},
  \bibinfo {author} {\bibfnamefont {J.~W.}\ \bibnamefont {{Kysar}}}, \ and\
  \bibinfo {author} {\bibfnamefont {J.}~\bibnamefont {{Hone}}},\ }\href
  {\doibase 10.1126/science.1157996} {\bibfield  {journal} {\bibinfo  {journal}
  {Science}\ }\textbf {\bibinfo {volume} {321}},\ \bibinfo {pages} {385}
  (\bibinfo {year} {2008})}\BibitemShut {NoStop}%
\bibitem [{\citenamefont {Balandin}\ \emph {et~al.}(2008)\citenamefont
  {Balandin}, \citenamefont {Ghosh}, \citenamefont {Bao}, \citenamefont
  {Calizo}, \citenamefont {Teweldebrhan}, \citenamefont {Miao},\ and\
  \citenamefont {Lau}}]{Balandin2008-902}%
  \BibitemOpen
  \bibfield  {author} {\bibinfo {author} {\bibfnamefont {A.~A.}\ \bibnamefont
  {Balandin}}, \bibinfo {author} {\bibfnamefont {S.}~\bibnamefont {Ghosh}},
  \bibinfo {author} {\bibfnamefont {W.}~\bibnamefont {Bao}}, \bibinfo {author}
  {\bibfnamefont {I.}~\bibnamefont {Calizo}}, \bibinfo {author} {\bibfnamefont
  {D.}~\bibnamefont {Teweldebrhan}}, \bibinfo {author} {\bibfnamefont
  {F.}~\bibnamefont {Miao}}, \ and\ \bibinfo {author} {\bibfnamefont {C.~N.}\
  \bibnamefont {Lau}},\ }\href@noop {} {\bibfield  {journal} {\bibinfo
  {journal} {Nano Lett.}\ }\textbf {\bibinfo {volume} {8}},\ \bibinfo {pages}
  {902} (\bibinfo {year} {2008})}\BibitemShut {NoStop}%
\bibitem [{\citenamefont {Novoselov}\ \emph {et~al.}(2004)\citenamefont
  {Novoselov}, \citenamefont {Geim}, \citenamefont {Morozov}, \citenamefont
  {Jiang}, \citenamefont {Katsnelson}, \citenamefont {Grigorieva},
  \citenamefont {Dubonos},\ and\ \citenamefont {Firsov}}]{Novoselov2004-666}%
  \BibitemOpen
  \bibfield  {author} {\bibinfo {author} {\bibfnamefont {K.~S.}\ \bibnamefont
  {Novoselov}}, \bibinfo {author} {\bibfnamefont {A.~K.}\ \bibnamefont {Geim}},
  \bibinfo {author} {\bibfnamefont {S.~V.}\ \bibnamefont {Morozov}}, \bibinfo
  {author} {\bibfnamefont {D.}~\bibnamefont {Jiang}}, \bibinfo {author}
  {\bibfnamefont {M.~I.}\ \bibnamefont {Katsnelson}}, \bibinfo {author}
  {\bibfnamefont {I.~V.}\ \bibnamefont {Grigorieva}}, \bibinfo {author}
  {\bibfnamefont {S.~V.}\ \bibnamefont {Dubonos}}, \ and\ \bibinfo {author}
  {\bibfnamefont {A.~A.}\ \bibnamefont {Firsov}},\ }\href@noop {} {\bibfield
  {journal} {\bibinfo  {journal} {Science}\ }\textbf {\bibinfo {volume}
  {306}},\ \bibinfo {pages} {666} (\bibinfo {year} {2004})}\BibitemShut
  {NoStop}%
\bibitem [{\citenamefont {{Boukhvalov}}\ and\ \citenamefont
  {{Katsnelson}}(2009)}]{Boukhvalov2009-344205}%
  \BibitemOpen
  \bibfield  {author} {\bibinfo {author} {\bibfnamefont {D.~W.}\ \bibnamefont
  {{Boukhvalov}}}\ and\ \bibinfo {author} {\bibfnamefont {M.~I.}\ \bibnamefont
  {{Katsnelson}}},\ }\href {\doibase 10.1088/0953-8984/21/34/344205} {\bibfield
   {journal} {\bibinfo  {journal} {Journal of Physics Condensed Matter}\
  }\textbf {\bibinfo {volume} {21}},\ \bibinfo {pages} {344205} (\bibinfo
  {year} {2009})}\BibitemShut {NoStop}%
\bibitem [{\citenamefont {Georgakilas}\ \emph {et~al.}(2012)\citenamefont
  {Georgakilas}, \citenamefont {Otyepka}, \citenamefont {Bourlinos},
  \citenamefont {Chandra}, \citenamefont {Kim}, \citenamefont {Christian~Kemp},
  \citenamefont {Hobza}, \citenamefont {Zboril},\ and\ \citenamefont
  {Kim}}]{Georgakilas2012-6156}%
  \BibitemOpen
  \bibfield  {author} {\bibinfo {author} {\bibfnamefont {V.}~\bibnamefont
  {Georgakilas}}, \bibinfo {author} {\bibfnamefont {M.}~\bibnamefont
  {Otyepka}}, \bibinfo {author} {\bibfnamefont {A.~B.}\ \bibnamefont
  {Bourlinos}}, \bibinfo {author} {\bibfnamefont {V.}~\bibnamefont {Chandra}},
  \bibinfo {author} {\bibfnamefont {N.}~\bibnamefont {Kim}}, \bibinfo {author}
  {\bibfnamefont {K.}~\bibnamefont {Christian~Kemp}}, \bibinfo {author}
  {\bibfnamefont {P.}~\bibnamefont {Hobza}}, \bibinfo {author} {\bibfnamefont
  {R.}~\bibnamefont {Zboril}}, \ and\ \bibinfo {author} {\bibfnamefont {K.~S.}\
  \bibnamefont {Kim}},\ }\href@noop {} {\bibfield  {journal} {\bibinfo
  {journal} {Chem. Rev.}\ }\textbf {\bibinfo {volume} {112}},\ \bibinfo {pages}
  {6156} (\bibinfo {year} {2012})}\BibitemShut {NoStop}%
\bibitem [{\citenamefont {Kuila}\ \emph {et~al.}(2012)\citenamefont {Kuila},
  \citenamefont {Bose}, \citenamefont {Mishra}, \citenamefont {Khanra},
  \citenamefont {Kim},\ and\ \citenamefont {Lee}}]{Kuila2012-1061}%
  \BibitemOpen
  \bibfield  {author} {\bibinfo {author} {\bibfnamefont {T.}~\bibnamefont
  {Kuila}}, \bibinfo {author} {\bibfnamefont {S.}~\bibnamefont {Bose}},
  \bibinfo {author} {\bibfnamefont {A.~K.}\ \bibnamefont {Mishra}}, \bibinfo
  {author} {\bibfnamefont {P.}~\bibnamefont {Khanra}}, \bibinfo {author}
  {\bibfnamefont {N.~H.}\ \bibnamefont {Kim}}, \ and\ \bibinfo {author}
  {\bibfnamefont {J.~H.}\ \bibnamefont {Lee}},\ }\href {\doibase
  http://dx.doi.org/10.1016/j.pmatsci.2012.03.002} {\bibfield  {journal}
  {\bibinfo  {journal} {Progress in Materials Science}\ }\textbf {\bibinfo
  {volume} {57}},\ \bibinfo {pages} {1061 } (\bibinfo {year}
  {2012})}\BibitemShut {NoStop}%
\bibitem [{\citenamefont {Boukhvalov}\ and\ \citenamefont
  {Katsnelson}(2008)}]{Boukhvalov2008-4373}%
  \BibitemOpen
  \bibfield  {author} {\bibinfo {author} {\bibfnamefont {D.~W.}\ \bibnamefont
  {Boukhvalov}}\ and\ \bibinfo {author} {\bibfnamefont {M.~I.}\ \bibnamefont
  {Katsnelson}},\ }\href@noop {} {\bibfield  {journal} {\bibinfo  {journal}
  {Nano Lett.}\ }\textbf {\bibinfo {volume} {8}},\ \bibinfo {pages} {4373}
  (\bibinfo {year} {2008})}\BibitemShut {NoStop}%
\bibitem [{\citenamefont {{Loh}}\ \emph {et~al.}(2010)\citenamefont {{Loh}},
  \citenamefont {{Bao}}, \citenamefont {{Eda}},\ and\ \citenamefont
  {{Chhowalla}}}]{Loh2010-1015}%
  \BibitemOpen
  \bibfield  {author} {\bibinfo {author} {\bibfnamefont {K.~P.}\ \bibnamefont
  {{Loh}}}, \bibinfo {author} {\bibfnamefont {Q.}~\bibnamefont {{Bao}}},
  \bibinfo {author} {\bibfnamefont {G.}~\bibnamefont {{Eda}}}, \ and\ \bibinfo
  {author} {\bibfnamefont {M.}~\bibnamefont {{Chhowalla}}},\ }\href {\doibase
  10.1038/nchem.907} {\bibfield  {journal} {\bibinfo  {journal} {Nature
  Chemistry}\ }\textbf {\bibinfo {volume} {2}},\ \bibinfo {pages} {1015}
  (\bibinfo {year} {2010})}\BibitemShut {NoStop}%
\bibitem [{\citenamefont {Englert}\ \emph {et~al.}(2011)\citenamefont
  {Englert}, \citenamefont {Dotzer}, \citenamefont {Yang}, \citenamefont
  {Schmid}, \citenamefont {Papp}, \citenamefont {Gottfried}, \citenamefont
  {Steinr\"uck}, \citenamefont {Spiecker}, \citenamefont {Hauke},\ and\
  \citenamefont {Hirsch}}]{Englert2011-279}%
  \BibitemOpen
  \bibfield  {author} {\bibinfo {author} {\bibfnamefont {J.~M.}\ \bibnamefont
  {Englert}}, \bibinfo {author} {\bibfnamefont {C.}~\bibnamefont {Dotzer}},
  \bibinfo {author} {\bibfnamefont {G.}~\bibnamefont {Yang}}, \bibinfo {author}
  {\bibfnamefont {M.}~\bibnamefont {Schmid}}, \bibinfo {author} {\bibfnamefont
  {C.}~\bibnamefont {Papp}}, \bibinfo {author} {\bibfnamefont {J.~M.}\
  \bibnamefont {Gottfried}}, \bibinfo {author} {\bibfnamefont {H.-P.}\
  \bibnamefont {Steinr\"uck}}, \bibinfo {author} {\bibfnamefont
  {E.}~\bibnamefont {Spiecker}}, \bibinfo {author} {\bibfnamefont
  {F.}~\bibnamefont {Hauke}}, \ and\ \bibinfo {author} {\bibfnamefont
  {A.}~\bibnamefont {Hirsch}},\ }\href@noop {} {\bibfield  {journal} {\bibinfo
  {journal} {Nature Chem.}\ }\textbf {\bibinfo {volume} {3}},\ \bibinfo {pages}
  {279} (\bibinfo {year} {2011})}\BibitemShut {NoStop}%
\bibitem [{\citenamefont {Cui}\ \emph {et~al.}(2011)\citenamefont {Cui},
  \citenamefont {Seo}, \citenamefont {Lee}, \citenamefont {Wang}, \citenamefont
  {Lee}, \citenamefont {Min},\ and\ \citenamefont {Lee}}]{Cui2011-6826}%
  \BibitemOpen
  \bibfield  {author} {\bibinfo {author} {\bibfnamefont {P.}~\bibnamefont
  {Cui}}, \bibinfo {author} {\bibfnamefont {S.}~\bibnamefont {Seo}}, \bibinfo
  {author} {\bibfnamefont {J.}~\bibnamefont {Lee}}, \bibinfo {author}
  {\bibfnamefont {L.}~\bibnamefont {Wang}}, \bibinfo {author} {\bibfnamefont
  {E.}~\bibnamefont {Lee}}, \bibinfo {author} {\bibfnamefont {M.}~\bibnamefont
  {Min}}, \ and\ \bibinfo {author} {\bibfnamefont {H.}~\bibnamefont {Lee}},\
  }\href@noop {} {\bibfield  {journal} {\bibinfo  {journal} {ACS Nano}\
  }\textbf {\bibinfo {volume} {5}},\ \bibinfo {pages} {6826} (\bibinfo {year}
  {2011})}\BibitemShut {NoStop}%
\bibitem [{\citenamefont {Kamat}(2010)}]{Kamat2010-520}%
  \BibitemOpen
  \bibfield  {author} {\bibinfo {author} {\bibfnamefont {P.~V.}\ \bibnamefont
  {Kamat}},\ }\href@noop {} {\bibfield  {journal} {\bibinfo  {journal} {J.
  Phys. Chem. Lett.}\ }\textbf {\bibinfo {volume} {1}},\ \bibinfo {pages} {520}
  (\bibinfo {year} {2010})}\BibitemShut {NoStop}%
\bibitem [{\citenamefont {Kamat}(2011)}]{Kamat2011-242}%
  \BibitemOpen
  \bibfield  {author} {\bibinfo {author} {\bibfnamefont {P.~V.}\ \bibnamefont
  {Kamat}},\ }\href@noop {} {\bibfield  {journal} {\bibinfo  {journal} {J.
  Phys. Chem. Lett.}\ }\textbf {\bibinfo {volume} {2}},\ \bibinfo {pages} {242}
  (\bibinfo {year} {2011})}\BibitemShut {NoStop}%
\bibitem [{\citenamefont {Zhou}\ and\ \citenamefont
  {Bongiorno}(2013)}]{Zhou2013-2484}%
  \BibitemOpen
  \bibfield  {author} {\bibinfo {author} {\bibfnamefont {S.}~\bibnamefont
  {Zhou}}\ and\ \bibinfo {author} {\bibfnamefont {A.}~\bibnamefont
  {Bongiorno}},\ }\href@noop {} {\bibfield  {journal} {\bibinfo  {journal}
  {Sci. Rep.}\ }\textbf {\bibinfo {volume} {3}},\ \bibinfo {pages} {2484}
  (\bibinfo {year} {2013})}\BibitemShut {NoStop}%
\bibitem [{\citenamefont {Dreyer}\ \emph {et~al.}(2010)\citenamefont {Dreyer},
  \citenamefont {Park}, \citenamefont {Bielawski},\ and\ \citenamefont
  {Ruoff}}]{Dreyer2010-228}%
  \BibitemOpen
  \bibfield  {author} {\bibinfo {author} {\bibfnamefont {D.~R.}\ \bibnamefont
  {Dreyer}}, \bibinfo {author} {\bibfnamefont {S.}~\bibnamefont {Park}},
  \bibinfo {author} {\bibfnamefont {C.~W.}\ \bibnamefont {Bielawski}}, \ and\
  \bibinfo {author} {\bibfnamefont {R.~S.}\ \bibnamefont {Ruoff}},\ }\href@noop
  {} {\bibfield  {journal} {\bibinfo  {journal} {Chemical Society reviews}\
  }\textbf {\bibinfo {volume} {39}},\ \bibinfo {pages} {228} (\bibinfo {year}
  {2010})}\BibitemShut {NoStop}%
\bibitem [{\citenamefont {Kim}\ \emph {et~al.}(2012)\citenamefont {Kim},
  \citenamefont {Zhou}, \citenamefont {Hu}, \citenamefont {Acik}, \citenamefont
  {Chabal}, \citenamefont {Berger}, \citenamefont {Heer}, \citenamefont
  {Bongiorno},\ and\ \citenamefont {Riedo}}]{Kim2012-544}%
  \BibitemOpen
  \bibfield  {author} {\bibinfo {author} {\bibfnamefont {S.}~\bibnamefont
  {Kim}}, \bibinfo {author} {\bibfnamefont {S.}~\bibnamefont {Zhou}}, \bibinfo
  {author} {\bibfnamefont {Y.~K.}\ \bibnamefont {Hu}}, \bibinfo {author}
  {\bibfnamefont {M.}~\bibnamefont {Acik}}, \bibinfo {author} {\bibfnamefont
  {Y.~J.}\ \bibnamefont {Chabal}}, \bibinfo {author} {\bibfnamefont
  {C.}~\bibnamefont {Berger}}, \bibinfo {author} {\bibfnamefont {W.~d.}\
  \bibnamefont {Heer}}, \bibinfo {author} {\bibfnamefont {A.}~\bibnamefont
  {Bongiorno}}, \ and\ \bibinfo {author} {\bibfnamefont {E.}~\bibnamefont
  {Riedo}},\ }\href {<Go to ISI>://WOS:000304320300022} {\bibfield  {journal}
  {\bibinfo  {journal} {Nature Materials}\ }\textbf {\bibinfo {volume} {11}},\
  \bibinfo {pages} {544} (\bibinfo {year} {2012})}\BibitemShut {NoStop}%
\bibitem [{\citenamefont {Pembroke}\ \emph {et~al.}(2013)\citenamefont
  {Pembroke}, \citenamefont {Ruan}, \citenamefont {Sinitskii}, \citenamefont
  {Corley}, \citenamefont {Yan}, \citenamefont {Sun},\ and\ \citenamefont
  {Tour}}]{Pembroke2013-138}%
  \BibitemOpen
  \bibfield  {author} {\bibinfo {author} {\bibfnamefont {E.}~\bibnamefont
  {Pembroke}}, \bibinfo {author} {\bibfnamefont {G.}~\bibnamefont {Ruan}},
  \bibinfo {author} {\bibfnamefont {A.}~\bibnamefont {Sinitskii}}, \bibinfo
  {author} {\bibfnamefont {D.~a.}\ \bibnamefont {Corley}}, \bibinfo {author}
  {\bibfnamefont {Z.}~\bibnamefont {Yan}}, \bibinfo {author} {\bibfnamefont
  {Z.}~\bibnamefont {Sun}}, \ and\ \bibinfo {author} {\bibfnamefont {J.~M.}\
  \bibnamefont {Tour}},\ }\href
  {http://link.springer.com/10.1007/s12274-013-0289-7} {\bibfield  {journal}
  {\bibinfo  {journal} {Nano Research}\ }\textbf {\bibinfo {volume} {6}},\
  \bibinfo {pages} {138} (\bibinfo {year} {2013})}\BibitemShut {NoStop}%
\bibitem [{\citenamefont {Robinson}\ \emph {et~al.}(2010)\citenamefont
  {Robinson}, \citenamefont {Burgess}, \citenamefont {Junkermeier},
  \citenamefont {Badescu}, \citenamefont {Reinecke}, \citenamefont {Perkins},
  \citenamefont {Zalalutdniov}, \citenamefont {Baldwin}, \citenamefont
  {Culbertson}, \citenamefont {Sheehan},\ and\ \citenamefont
  {et~al.}}]{Robinson2010-3001}%
  \BibitemOpen
  \bibfield  {author} {\bibinfo {author} {\bibfnamefont {J.~T.}\ \bibnamefont
  {Robinson}}, \bibinfo {author} {\bibfnamefont {J.~S.}\ \bibnamefont
  {Burgess}}, \bibinfo {author} {\bibfnamefont {C.~E.}\ \bibnamefont
  {Junkermeier}}, \bibinfo {author} {\bibfnamefont {S.~C.}\ \bibnamefont
  {Badescu}}, \bibinfo {author} {\bibfnamefont {T.~L.}\ \bibnamefont
  {Reinecke}}, \bibinfo {author} {\bibfnamefont {F.~K.}\ \bibnamefont
  {Perkins}}, \bibinfo {author} {\bibfnamefont {M.~K.}\ \bibnamefont
  {Zalalutdniov}}, \bibinfo {author} {\bibfnamefont {J.~W.}\ \bibnamefont
  {Baldwin}}, \bibinfo {author} {\bibfnamefont {J.~C.}\ \bibnamefont
  {Culbertson}}, \bibinfo {author} {\bibfnamefont {P.~E.}\ \bibnamefont
  {Sheehan}}, \ and\ \bibinfo {author} {\bibnamefont {et~al.}},\ }\href@noop {}
  {\bibfield  {journal} {\bibinfo  {journal} {Nano letters}\ }\textbf {\bibinfo
  {volume} {10}},\ \bibinfo {pages} {3001} (\bibinfo {year}
  {2010})}\BibitemShut {NoStop}%
\bibitem [{\citenamefont {Abanin}\ \emph {et~al.}(2010)\citenamefont {Abanin},
  \citenamefont {Shytov},\ and\ \citenamefont {Levitov}}]{Abanin2010-086802}%
  \BibitemOpen
  \bibfield  {author} {\bibinfo {author} {\bibfnamefont {D.~A.}\ \bibnamefont
  {Abanin}}, \bibinfo {author} {\bibfnamefont {A.~V.}\ \bibnamefont {Shytov}},
  \ and\ \bibinfo {author} {\bibfnamefont {L.~S.}\ \bibnamefont {Levitov}},\
  }\href@noop {} {\bibfield  {journal} {\bibinfo  {journal} {Phys. Rev. Lett.}\
  }\textbf {\bibinfo {volume} {105}},\ \bibinfo {pages} {086802} (\bibinfo
  {year} {2010})}\BibitemShut {NoStop}%
\bibitem [{\citenamefont {Suarez}\ \emph {et~al.}(2011)\citenamefont {Suarez},
  \citenamefont {Radovic}, \citenamefont {Bar-Ziv},\ and\ \citenamefont
  {Sofo}}]{Suarez2011-146802}%
  \BibitemOpen
  \bibfield  {author} {\bibinfo {author} {\bibfnamefont {A.~M.}\ \bibnamefont
  {Suarez}}, \bibinfo {author} {\bibfnamefont {L.~R.}\ \bibnamefont {Radovic}},
  \bibinfo {author} {\bibfnamefont {E.}~\bibnamefont {Bar-Ziv}}, \ and\
  \bibinfo {author} {\bibfnamefont {J.~O.}\ \bibnamefont {Sofo}},\ }\href@noop
  {} {\bibfield  {journal} {\bibinfo  {journal} {Phys. Rev. Lett.}\ }\textbf
  {\bibinfo {volume} {106}},\ \bibinfo {pages} {146802} (\bibinfo {year}
  {2011})}\BibitemShut {NoStop}%
\bibitem [{\citenamefont {Solenov}\ and\ \citenamefont
  {Velizhanin}(2012)}]{Solenov2012-095504}%
  \BibitemOpen
  \bibfield  {author} {\bibinfo {author} {\bibfnamefont {D.}~\bibnamefont
  {Solenov}}\ and\ \bibinfo {author} {\bibfnamefont {K.~A.}\ \bibnamefont
  {Velizhanin}},\ }\href@noop {} {\bibfield  {journal} {\bibinfo  {journal}
  {Phys. Rev. Lett.}\ }\textbf {\bibinfo {volume} {109}},\ \bibinfo {pages}
  {095504} (\bibinfo {year} {2012})}\BibitemShut {NoStop}%
\bibitem [{\citenamefont {Sorbello}(1997)}]{Sorbello1997-159}%
  \BibitemOpen
  \bibfield  {author} {\bibinfo {author} {\bibfnamefont {R.~S.}\ \bibnamefont
  {Sorbello}},\ }\href@noop {} {\bibfield  {journal} {\bibinfo  {journal}
  {Solid State Phys.}\ }\textbf {\bibinfo {volume} {51}},\ \bibinfo {pages}
  {159} (\bibinfo {year} {1997})}\BibitemShut {NoStop}%
\bibitem [{\citenamefont {Yang}\ \emph {et~al.}(2008)\citenamefont {Yang},
  \citenamefont {Berber}, \citenamefont {Liu}, \citenamefont {Miller},\ and\
  \citenamefont {Tom\'anek}}]{Yang2008-124709}%
  \BibitemOpen
  \bibfield  {author} {\bibinfo {author} {\bibfnamefont {T.}~\bibnamefont
  {Yang}}, \bibinfo {author} {\bibfnamefont {S.}~\bibnamefont {Berber}},
  \bibinfo {author} {\bibfnamefont {J.-F.}\ \bibnamefont {Liu}}, \bibinfo
  {author} {\bibfnamefont {G.~P.}\ \bibnamefont {Miller}}, \ and\ \bibinfo
  {author} {\bibfnamefont {D.}~\bibnamefont {Tom\'anek}},\ }\href@noop {}
  {\bibfield  {journal} {\bibinfo  {journal} {J. Chem. Phys.}\ }\textbf
  {\bibinfo {volume} {128}},\ \bibinfo {pages} {124709} (\bibinfo {year}
  {2008})}\BibitemShut {NoStop}%
\bibitem [{\citenamefont {Voloshina}\ \emph {et~al.}(2011)\citenamefont
  {Voloshina}, \citenamefont {Mollenhauer}, \citenamefont {Chiappisi},\ and\
  \citenamefont {Paulus}}]{Voloshina2011-220}%
  \BibitemOpen
  \bibfield  {author} {\bibinfo {author} {\bibfnamefont {E.~N.}\ \bibnamefont
  {Voloshina}}, \bibinfo {author} {\bibfnamefont {D.}~\bibnamefont
  {Mollenhauer}}, \bibinfo {author} {\bibfnamefont {L.}~\bibnamefont
  {Chiappisi}}, \ and\ \bibinfo {author} {\bibfnamefont {B.}~\bibnamefont
  {Paulus}},\ }\href@noop {} {\bibfield  {journal} {\bibinfo  {journal} {Chem.
  Phys. Lett.}\ }\textbf {\bibinfo {volume} {510}},\ \bibinfo {pages} {220}
  (\bibinfo {year} {2011})}\BibitemShut {NoStop}%
\bibitem [{\citenamefont {Hellmann}(1937)}]{Hellmann1937}%
  \BibitemOpen
  \bibfield  {author} {\bibinfo {author} {\bibfnamefont {H.}~\bibnamefont
  {Hellmann}},\ }\href@noop {} {\emph {\bibinfo {title} {Einf\"uhrung in die
  Quantenchemie}}}\ (\bibinfo  {publisher} {Deuticke},\ \bibinfo {address}
  {Leipzig},\ \bibinfo {year} {1937})\BibitemShut {NoStop}%
\bibitem [{\citenamefont {Feynman}(1939)}]{Feynman1939-340}%
  \BibitemOpen
  \bibfield  {author} {\bibinfo {author} {\bibfnamefont {R.~P.}\ \bibnamefont
  {Feynman}},\ }\href@noop {} {\bibfield  {journal} {\bibinfo  {journal} {Phys.
  Rev.}\ }\textbf {\bibinfo {volume} {56}},\ \bibinfo {pages} {340} (\bibinfo
  {year} {1939})}\BibitemShut {NoStop}%
\bibitem [{\citenamefont {Economou}(1990)}]{Economou1990}%
  \BibitemOpen
  \bibfield  {author} {\bibinfo {author} {\bibfnamefont {E.~N.}\ \bibnamefont
  {Economou}},\ }\href@noop {} {\emph {\bibinfo {title} {Green's Functions in
  Quantum Physics}}},\ \bibinfo {edition} {2nd}\ ed.\ (\bibinfo  {publisher}
  {Springer-Verlag},\ \bibinfo {address} {New York},\ \bibinfo {year}
  {1990})\BibitemShut {NoStop}%
\bibitem [{Note1()}]{Note1}%
  \BibitemOpen
  \bibinfo {note} {The ambiguity in defining the electron wind force locally is
  reminiscent of gauge freedom in electrodynamics. Indeed, one can add a
  gradient of an arbitrary scalar function to the wind force defined locally.
  If the chosen scalar function is periodic over the substrate unit cell, its
  specific form does not affect the rigorous definition of the wind force,
  i.e., the one done via averaging.}\BibitemShut {Stop}%
\bibitem [{\citenamefont {Giuliani}\ and\ \citenamefont
  {Vignale}(2005)}]{Giuliani2005}%
  \BibitemOpen
  \bibfield  {author} {\bibinfo {author} {\bibfnamefont {G.~F.}\ \bibnamefont
  {Giuliani}}\ and\ \bibinfo {author} {\bibfnamefont {G.}~\bibnamefont
  {Vignale}},\ }\href@noop {} {\emph {\bibinfo {title} {Quantum Theory of the
  Electron Liquid}}},\ \bibinfo {edition} {1st}\ ed.\ (\bibinfo  {publisher}
  {Cambridge University Press},\ \bibinfo {address} {Cambridge},\ \bibinfo
  {year} {2005})\BibitemShut {NoStop}%
\bibitem [{\citenamefont {van Schilfgaarde}\ and\ \citenamefont
  {Katsnelson}(2011)}]{Schilfgaarde2011-081409}%
  \BibitemOpen
  \bibfield  {author} {\bibinfo {author} {\bibfnamefont {M.}~\bibnamefont {van
  Schilfgaarde}}\ and\ \bibinfo {author} {\bibfnamefont {M.~I.}\ \bibnamefont
  {Katsnelson}},\ }\href@noop {} {\bibfield  {journal} {\bibinfo  {journal}
  {Physical Review B}\ }\textbf {\bibinfo {volume} {83}},\ \bibinfo {pages}
  {081409(R)} (\bibinfo {year} {2011})}\BibitemShut {NoStop}%
\bibitem [{\citenamefont {Mitchell}\ and\ \citenamefont
  {Wallis}(1966)}]{Mitchell1966-581}%
  \BibitemOpen
  \bibfield  {author} {\bibinfo {author} {\bibfnamefont {D.~L.}\ \bibnamefont
  {Mitchell}}\ and\ \bibinfo {author} {\bibfnamefont {R.~F.}\ \bibnamefont
  {Wallis}},\ }\href@noop {} {\bibfield  {journal} {\bibinfo  {journal} {Phys.
  Rev.}\ }\textbf {\bibinfo {volume} {151}},\ \bibinfo {pages} {581} (\bibinfo
  {year} {1966})}\BibitemShut {NoStop}%
\bibitem [{\citenamefont {Carter}\ and\ \citenamefont
  {Bates}(1971)}]{Dimmock1971}%
  \BibitemOpen
  \bibinfo {editor} {\bibfnamefont {D.~L.}\ \bibnamefont {Carter}}\ and\
  \bibinfo {editor} {\bibfnamefont {R.~T.}\ \bibnamefont {Bates}},\ eds.,\
  \enquote {\bibinfo {title} {The physics of semimetals and narrow gap
  semiconductors},}\ \ (\bibinfo  {publisher} {Pergamon},\ \bibinfo {address}
  {Oxford},\ \bibinfo {year} {1971})\BibitemShut {NoStop}%
\bibitem [{\citenamefont {Kang}\ and\ \citenamefont
  {Wise}(1997)}]{Kang1997-1632}%
  \BibitemOpen
  \bibfield  {author} {\bibinfo {author} {\bibfnamefont {I.}~\bibnamefont
  {Kang}}\ and\ \bibinfo {author} {\bibfnamefont {F.~W.}\ \bibnamefont
  {Wise}},\ }\href@noop {} {\bibfield  {journal} {\bibinfo  {journal} {J. Opt.
  Soc. Am. B}\ }\textbf {\bibinfo {volume} {14}},\ \bibinfo {pages} {1632}
  (\bibinfo {year} {1997})}\BibitemShut {NoStop}%
\bibitem [{\citenamefont {Semenoff}(1984)}]{Semenoff1984-2449}%
  \BibitemOpen
  \bibfield  {author} {\bibinfo {author} {\bibfnamefont {G.~W.}\ \bibnamefont
  {Semenoff}},\ }\href@noop {} {\bibfield  {journal} {\bibinfo  {journal}
  {Phys. Rev. Lett.}\ }\textbf {\bibinfo {volume} {53}},\ \bibinfo {pages}
  {2449} (\bibinfo {year} {1984})}\BibitemShut {NoStop}%
\bibitem [{\citenamefont {Neto}\ \emph {et~al.}(2009)\citenamefont {Neto},
  \citenamefont {Guinea}, \citenamefont {Peres}, \citenamefont {Novoselov},\
  and\ \citenamefont {Geim}}]{CastroNeto2009-109}%
  \BibitemOpen
  \bibfield  {author} {\bibinfo {author} {\bibfnamefont {A.~H.~C.}\
  \bibnamefont {Neto}}, \bibinfo {author} {\bibfnamefont {F.}~\bibnamefont
  {Guinea}}, \bibinfo {author} {\bibfnamefont {N.~M.~R.}\ \bibnamefont
  {Peres}}, \bibinfo {author} {\bibfnamefont {K.~S.}\ \bibnamefont
  {Novoselov}}, \ and\ \bibinfo {author} {\bibfnamefont {A.~K.}\ \bibnamefont
  {Geim}},\ }\href@noop {} {\bibfield  {journal} {\bibinfo  {journal} {Rev.
  Mod. Phys.}\ }\textbf {\bibinfo {volume} {81}},\ \bibinfo {pages} {109}
  (\bibinfo {year} {2009})}\BibitemShut {NoStop}%
\bibitem [{\citenamefont {Das\:Sarma}\ \emph {et~al.}(2011)\citenamefont
  {Das\:Sarma}, \citenamefont {Adam}, \citenamefont {Hwang},\ and\
  \citenamefont {Rossi}}]{DasSarma2011-407}%
  \BibitemOpen
  \bibfield  {author} {\bibinfo {author} {\bibfnamefont {S.}~\bibnamefont
  {Das\:Sarma}}, \bibinfo {author} {\bibfnamefont {S.}~\bibnamefont {Adam}},
  \bibinfo {author} {\bibfnamefont {E.~H.}\ \bibnamefont {Hwang}}, \ and\
  \bibinfo {author} {\bibfnamefont {E.}~\bibnamefont {Rossi}},\ }\href@noop {}
  {\bibfield  {journal} {\bibinfo  {journal} {Rev. Mod. Phys.}\ }\textbf
  {\bibinfo {volume} {83}},\ \bibinfo {pages} {407} (\bibinfo {year}
  {2011})}\BibitemShut {NoStop}%
\bibitem [{\citenamefont {Ponomarenko}\ \emph {et~al.}(2009)\citenamefont
  {Ponomarenko}, \citenamefont {Yang}, \citenamefont {Mohiuddin}, \citenamefont
  {Katsnelson}, \citenamefont {Novoselov}, \citenamefont {Morozov},
  \citenamefont {Zhukov}, \citenamefont {Schedin}, \citenamefont {Hill},\ and\
  \citenamefont {Geim}}]{Ponomarenko2009-206603}%
  \BibitemOpen
  \bibfield  {author} {\bibinfo {author} {\bibfnamefont {L.~A.}\ \bibnamefont
  {Ponomarenko}}, \bibinfo {author} {\bibfnamefont {R.}~\bibnamefont {Yang}},
  \bibinfo {author} {\bibfnamefont {T.~M.}\ \bibnamefont {Mohiuddin}}, \bibinfo
  {author} {\bibfnamefont {M.~I.}\ \bibnamefont {Katsnelson}}, \bibinfo
  {author} {\bibfnamefont {K.~S.}\ \bibnamefont {Novoselov}}, \bibinfo {author}
  {\bibfnamefont {S.~V.}\ \bibnamefont {Morozov}}, \bibinfo {author}
  {\bibfnamefont {A.~A.}\ \bibnamefont {Zhukov}}, \bibinfo {author}
  {\bibfnamefont {F.}~\bibnamefont {Schedin}}, \bibinfo {author} {\bibfnamefont
  {E.~W.}\ \bibnamefont {Hill}}, \ and\ \bibinfo {author} {\bibfnamefont
  {A.~K.}\ \bibnamefont {Geim}},\ }\href@noop {} {\bibfield  {journal}
  {\bibinfo  {journal} {Phys. Rev. Lett.}\ }\textbf {\bibinfo {volume} {102}},\
  \bibinfo {pages} {206603} (\bibinfo {year} {2009})}\BibitemShut {NoStop}%
\bibitem [{\citenamefont {Shytov}\ \emph
  {et~al.}(2007{\natexlab{a}})\citenamefont {Shytov}, \citenamefont
  {Katsnelson},\ and\ \citenamefont {Levitov}}]{Shytov2007-236801}%
  \BibitemOpen
  \bibfield  {author} {\bibinfo {author} {\bibfnamefont {A.~V.}\ \bibnamefont
  {Shytov}}, \bibinfo {author} {\bibfnamefont {M.~I.}\ \bibnamefont
  {Katsnelson}}, \ and\ \bibinfo {author} {\bibfnamefont {L.~S.}\ \bibnamefont
  {Levitov}},\ }\href@noop {} {\bibfield  {journal} {\bibinfo  {journal} {Phys.
  Rev. Lett.}\ }\textbf {\bibinfo {volume} {99}},\ \bibinfo {pages} {236801}
  (\bibinfo {year} {2007}{\natexlab{a}})}\BibitemShut {NoStop}%
\bibitem [{\citenamefont {Biswas}\ \emph {et~al.}(2007)\citenamefont {Biswas},
  \citenamefont {Sachdev},\ and\ \citenamefont {Son}}]{Biswas2007-205122}%
  \BibitemOpen
  \bibfield  {author} {\bibinfo {author} {\bibfnamefont {R.~R.}\ \bibnamefont
  {Biswas}}, \bibinfo {author} {\bibfnamefont {S.}~\bibnamefont {Sachdev}}, \
  and\ \bibinfo {author} {\bibfnamefont {D.~T.}\ \bibnamefont {Son}},\
  }\href@noop {} {\bibfield  {journal} {\bibinfo  {journal} {Physical Review
  B}\ }\textbf {\bibinfo {volume} {76}},\ \bibinfo {pages} {205122} (\bibinfo
  {year} {2007})}\BibitemShut {NoStop}%
\bibitem [{\citenamefont {Shytov}\ \emph
  {et~al.}(2007{\natexlab{b}})\citenamefont {Shytov}, \citenamefont
  {Katsnelson},\ and\ \citenamefont {Levitov}}]{Shytov2007-246802}%
  \BibitemOpen
  \bibfield  {author} {\bibinfo {author} {\bibfnamefont {A.~V.}\ \bibnamefont
  {Shytov}}, \bibinfo {author} {\bibfnamefont {M.~I.}\ \bibnamefont
  {Katsnelson}}, \ and\ \bibinfo {author} {\bibfnamefont {L.~S.}\ \bibnamefont
  {Levitov}},\ }\href@noop {} {\bibfield  {journal} {\bibinfo  {journal} {Phys.
  Rev. Lett.}\ }\textbf {\bibinfo {volume} {99}},\ \bibinfo {pages} {246802}
  (\bibinfo {year} {2007}{\natexlab{b}})}\BibitemShut {NoStop}%
\bibitem [{\citenamefont {Terekhov}\ \emph {et~al.}(2008)\citenamefont
  {Terekhov}, \citenamefont {Milstein}, \citenamefont {Kotov},\ and\
  \citenamefont {Sushkov}}]{Terekhov2008-076803}%
  \BibitemOpen
  \bibfield  {author} {\bibinfo {author} {\bibfnamefont {I.~S.}\ \bibnamefont
  {Terekhov}}, \bibinfo {author} {\bibfnamefont {A.~I.}\ \bibnamefont
  {Milstein}}, \bibinfo {author} {\bibfnamefont {V.~N.}\ \bibnamefont {Kotov}},
  \ and\ \bibinfo {author} {\bibfnamefont {O.~P.}\ \bibnamefont {Sushkov}},\
  }\href@noop {} {\bibfield  {journal} {\bibinfo  {journal} {Phys. Rev. Lett.}\
  }\textbf {\bibinfo {volume} {100}},\ \bibinfo {pages} {076803} (\bibinfo
  {year} {2008})}\BibitemShut {NoStop}%
\bibitem [{\citenamefont {Solenov}\ \emph {et~al.}(2013)\citenamefont
  {Solenov}, \citenamefont {Junkermeier}, \citenamefont {Reinecke},\ and\
  \citenamefont {Velizhanin}}]{Solenov2013-115502}%
  \BibitemOpen
  \bibfield  {author} {\bibinfo {author} {\bibfnamefont {D.}~\bibnamefont
  {Solenov}}, \bibinfo {author} {\bibfnamefont {C.}~\bibnamefont
  {Junkermeier}}, \bibinfo {author} {\bibfnamefont {T.~L.}\ \bibnamefont
  {Reinecke}}, \ and\ \bibinfo {author} {\bibfnamefont {K.~A.}\ \bibnamefont
  {Velizhanin}},\ }\href@noop {} {\bibfield  {journal} {\bibinfo  {journal}
  {Phys. Rev. Lett.}\ }\textbf {\bibinfo {volume} {111}},\ \bibinfo {pages}
  {115502} (\bibinfo {year} {2013})}\BibitemShut {NoStop}%
\bibitem [{\citenamefont {Ashcroft}\ and\ \citenamefont
  {Mermin}(1976)}]{Ashcroft1976}%
  \BibitemOpen
  \bibfield  {author} {\bibinfo {author} {\bibfnamefont {N.~W.}\ \bibnamefont
  {Ashcroft}}\ and\ \bibinfo {author} {\bibfnamefont {N.~D.}\ \bibnamefont
  {Mermin}},\ }\href@noop {} {\emph {\bibinfo {title} {Solid State Physics}}}\
  (\bibinfo  {publisher} {Holt, Rinehart and Winston},\ \bibinfo {address} {New
  York},\ \bibinfo {year} {1976})\BibitemShut {NoStop}%
\bibitem [{\citenamefont {Kresse}\ and\ \citenamefont
  {Furthmuller}(1996{\natexlab{a}})}]{Kresse1996-15}%
  \BibitemOpen
  \bibfield  {author} {\bibinfo {author} {\bibfnamefont {G.}~\bibnamefont
  {Kresse}}\ and\ \bibinfo {author} {\bibfnamefont {J.}~\bibnamefont
  {Furthmuller}},\ }\href@noop {} {\bibfield  {journal} {\bibinfo  {journal}
  {Comp. Mat. Science}\ }\textbf {\bibinfo {volume} {6}},\ \bibinfo {pages}
  {15} (\bibinfo {year} {1996}{\natexlab{a}})}\BibitemShut {NoStop}%
\bibitem [{\citenamefont {Kresse}\ and\ \citenamefont
  {Furthmuller}(1996{\natexlab{b}})}]{Kresse1996-11169}%
  \BibitemOpen
  \bibfield  {author} {\bibinfo {author} {\bibfnamefont {G.}~\bibnamefont
  {Kresse}}\ and\ \bibinfo {author} {\bibfnamefont {J.}~\bibnamefont
  {Furthmuller}},\ }\href@noop {} {\bibfield  {journal} {\bibinfo  {journal}
  {Phys. Rev. B}\ }\textbf {\bibinfo {volume} {54}},\ \bibinfo {pages} {11169}
  (\bibinfo {year} {1996}{\natexlab{b}})}\BibitemShut {NoStop}%
\bibitem [{\citenamefont {Perdew}\ \emph {et~al.}(1996)\citenamefont {Perdew},
  \citenamefont {Burke},\ and\ \citenamefont {M.~Ernzerhof}}]{Perdew1996-3865}%
  \BibitemOpen
  \bibfield  {author} {\bibinfo {author} {\bibfnamefont {J.~P.}\ \bibnamefont
  {Perdew}}, \bibinfo {author} {\bibfnamefont {K.}~\bibnamefont {Burke}}, \
  and\ \bibinfo {author} {\bibfnamefont {M.}~\bibnamefont {M.~Ernzerhof}},\
  }\href@noop {} {\bibfield  {journal} {\bibinfo  {journal} {Phys. Rev. Lett.}\
  }\textbf {\bibinfo {volume} {77}},\ \bibinfo {pages} {3865} (\bibinfo {year}
  {1996})}\BibitemShut {NoStop}%
\bibitem [{\citenamefont {Perdew}\ \emph {et~al.}(1997)\citenamefont {Perdew},
  \citenamefont {Burke},\ and\ \citenamefont {M.~Ernzerhof}}]{Perdew1997-1396}%
  \BibitemOpen
  \bibfield  {author} {\bibinfo {author} {\bibfnamefont {J.~P.}\ \bibnamefont
  {Perdew}}, \bibinfo {author} {\bibfnamefont {K.}~\bibnamefont {Burke}}, \
  and\ \bibinfo {author} {\bibfnamefont {M.}~\bibnamefont {M.~Ernzerhof}},\
  }\href@noop {} {\bibfield  {journal} {\bibinfo  {journal} {Phys. Rev. Lett.}\
  }\textbf {\bibinfo {volume} {78}},\ \bibinfo {pages} {1396} (\bibinfo {year}
  {1997})}\BibitemShut {NoStop}%
\bibitem [{\citenamefont {Blochl}(1994)}]{Blochl1994-17953}%
  \BibitemOpen
  \bibfield  {author} {\bibinfo {author} {\bibfnamefont {P.~E.}\ \bibnamefont
  {Blochl}},\ }\href@noop {} {\bibfield  {journal} {\bibinfo  {journal} {Phys.
  Rev. B}\ }\textbf {\bibinfo {volume} {50}},\ \bibinfo {pages} {17953}
  (\bibinfo {year} {1994})}\BibitemShut {NoStop}%
\bibitem [{\citenamefont {Kresse}\ and\ \citenamefont
  {Joubert}(1999)}]{Kresse1999-1758}%
  \BibitemOpen
  \bibfield  {author} {\bibinfo {author} {\bibfnamefont {G.}~\bibnamefont
  {Kresse}}\ and\ \bibinfo {author} {\bibfnamefont {D.}~\bibnamefont
  {Joubert}},\ }\href@noop {} {\bibfield  {journal} {\bibinfo  {journal} {Phys.
  Rev. B}\ }\textbf {\bibinfo {volume} {59}},\ \bibinfo {pages} {1758}
  (\bibinfo {year} {1999})}\BibitemShut {NoStop}%
\bibitem [{\citenamefont {Topsakal}\ \emph {et~al.}(2013)\citenamefont
  {Topsakal}, \citenamefont {Gurel},\ and\ \citenamefont
  {Ciraci}}]{Topsakal2013-5943}%
  \BibitemOpen
  \bibfield  {author} {\bibinfo {author} {\bibfnamefont {M.}~\bibnamefont
  {Topsakal}}, \bibinfo {author} {\bibfnamefont {H.}~\bibnamefont {Gurel}}, \
  and\ \bibinfo {author} {\bibfnamefont {S.}~\bibnamefont {Ciraci}},\
  }\href@noop {} {\bibfield  {journal} {\bibinfo  {journal} {J. Phys. Chem. C}\
  }\textbf {\bibinfo {volume} {117}},\ \bibinfo {pages} {5943} (\bibinfo {year}
  {2013})}\BibitemShut {NoStop}%
\bibitem [{\citenamefont {Neugebauer}\ and\ \citenamefont
  {Scheffler}(1992)}]{Neugebauer1992-16067}%
  \BibitemOpen
  \bibfield  {author} {\bibinfo {author} {\bibfnamefont {J.}~\bibnamefont
  {Neugebauer}}\ and\ \bibinfo {author} {\bibfnamefont {M.}~\bibnamefont
  {Scheffler}},\ }\href@noop {} {\bibfield  {journal} {\bibinfo  {journal}
  {Phys. Rev. B}\ }\textbf {\bibinfo {volume} {46}},\ \bibinfo {pages} {16067}
  (\bibinfo {year} {1992})}\BibitemShut {NoStop}%
\bibitem [{\citenamefont {Makov}\ and\ \citenamefont
  {Payne}(1995)}]{Makov1995-4014}%
  \BibitemOpen
  \bibfield  {author} {\bibinfo {author} {\bibfnamefont {G.}~\bibnamefont
  {Makov}}\ and\ \bibinfo {author} {\bibfnamefont {M.~C.}\ \bibnamefont
  {Payne}},\ }\href@noop {} {\bibfield  {journal} {\bibinfo  {journal} {Phys.
  Rev. B}\ }\textbf {\bibinfo {volume} {51}},\ \bibinfo {pages} {4014}
  (\bibinfo {year} {1995})}\BibitemShut {NoStop}%
\bibitem [{\citenamefont {Monkhorst}\ and\ \citenamefont
  {Pack}(1976)}]{Monkhorst1976-5188}%
  \BibitemOpen
  \bibfield  {author} {\bibinfo {author} {\bibfnamefont {H.~J.}\ \bibnamefont
  {Monkhorst}}\ and\ \bibinfo {author} {\bibfnamefont {J.~D.}\ \bibnamefont
  {Pack}},\ }\href@noop {} {\bibfield  {journal} {\bibinfo  {journal} {Phys.
  Rev. B}\ }\textbf {\bibinfo {volume} {13}},\ \bibinfo {pages} {5188}
  (\bibinfo {year} {1976})}\BibitemShut {NoStop}%
\bibitem [{\citenamefont {Sheppard}\ \emph {et~al.}(2008)\citenamefont
  {Sheppard}, \citenamefont {Terrell},\ and\ \citenamefont
  {Henkelman}}]{Sheppard2008-134106}%
  \BibitemOpen
  \bibfield  {author} {\bibinfo {author} {\bibfnamefont {D.}~\bibnamefont
  {Sheppard}}, \bibinfo {author} {\bibfnamefont {R.}~\bibnamefont {Terrell}}, \
  and\ \bibinfo {author} {\bibfnamefont {G.}~\bibnamefont {Henkelman}},\
  }\href@noop {} {\bibfield  {journal} {\bibinfo  {journal} {J. Chem. Phys.}\
  }\textbf {\bibinfo {volume} {128}},\ \bibinfo {pages} {134106} (\bibinfo
  {year} {2008})}\BibitemShut {NoStop}%
\bibitem [{\citenamefont {Henkelman}\ \emph {et~al.}(2006)\citenamefont
  {Henkelman}, \citenamefont {Arnaldsson},\ and\ \citenamefont
  {Jonsson}}]{Henkelman2006-254}%
  \BibitemOpen
  \bibfield  {author} {\bibinfo {author} {\bibfnamefont {G.}~\bibnamefont
  {Henkelman}}, \bibinfo {author} {\bibfnamefont {A.}~\bibnamefont
  {Arnaldsson}}, \ and\ \bibinfo {author} {\bibfnamefont {H.}~\bibnamefont
  {Jonsson}},\ }\href@noop {} {\bibfield  {journal} {\bibinfo  {journal}
  {Comput. Mater. Sci.}\ }\textbf {\bibinfo {volume} {36}},\ \bibinfo {pages}
  {254} (\bibinfo {year} {2006})}\BibitemShut {NoStop}%
\bibitem [{Bad()}]{Bader_scripts}%
  \BibitemOpen
  \href@noop {} {}\bibinfo {note}
  {Http://theory.cm.utexas.edu/bader}\BibitemShut {NoStop}%
\bibitem [{\citenamefont {Vineyard}(1957)}]{Vineyard1957-121}%
  \BibitemOpen
  \bibfield  {author} {\bibinfo {author} {\bibfnamefont {G.~H.}\ \bibnamefont
  {Vineyard}},\ }\href@noop {} {\bibfield  {journal} {\bibinfo  {journal} {J.
  Phys. Chem. Solids}\ }\textbf {\bibinfo {volume} {3}},\ \bibinfo {pages}
  {121} (\bibinfo {year} {1957})}\BibitemShut {NoStop}%
\bibitem [{\citenamefont {Henkelman}()}]{VTST_scripts}%
  \BibitemOpen
  \bibfield  {author} {\bibinfo {author} {\bibfnamefont {G.}~\bibnamefont
  {Henkelman}},\ }\href@noop {} {\enquote {\bibinfo {title} {{VASP TST
  Tools}},}\ }\bibinfo {note}
  {Http://theory.cm.utexas.edu/vtsttools/dynmat/}\BibitemShut {NoStop}%
\bibitem [{\citenamefont {Haenggi}\ \emph {et~al.}(1990)\citenamefont
  {Haenggi}, \citenamefont {Talkner},\ and\ \citenamefont
  {Borkovec}}]{Haenggi1990-251}%
  \BibitemOpen
  \bibfield  {author} {\bibinfo {author} {\bibfnamefont {P.}~\bibnamefont
  {Haenggi}}, \bibinfo {author} {\bibfnamefont {P.}~\bibnamefont {Talkner}}, \
  and\ \bibinfo {author} {\bibfnamefont {M.}~\bibnamefont {Borkovec}},\
  }\href@noop {} {\bibfield  {journal} {\bibinfo  {journal} {Rev. Mod. Phys.}\
  }\textbf {\bibinfo {volume} {62}},\ \bibinfo {pages} {251} (\bibinfo {year}
  {1990})}\BibitemShut {NoStop}%
\bibitem [{\citenamefont {Park}\ \emph {et~al.}(2009)\citenamefont {Park},
  \citenamefont {Dikin}, \citenamefont {Nguyen},\ and\ \citenamefont
  {Ruoff}}]{Park2009-15801}%
  \BibitemOpen
  \bibfield  {author} {\bibinfo {author} {\bibfnamefont {S.}~\bibnamefont
  {Park}}, \bibinfo {author} {\bibfnamefont {D.~A.}\ \bibnamefont {Dikin}},
  \bibinfo {author} {\bibfnamefont {S.~T.}\ \bibnamefont {Nguyen}}, \ and\
  \bibinfo {author} {\bibfnamefont {R.~S.}\ \bibnamefont {Ruoff}},\ }\href@noop
  {} {\bibfield  {journal} {\bibinfo  {journal} {J. Phys. Chem. C. Lett.}\
  }\textbf {\bibinfo {volume} {113}},\ \bibinfo {pages} {15801} (\bibinfo
  {year} {2009})}\BibitemShut {NoStop}%
\bibitem [{\citenamefont {Zhang}\ \emph {et~al.}(2010)\citenamefont {Zhang},
  \citenamefont {Yang}, \citenamefont {Shen}, \citenamefont {Cheng},
  \citenamefont {Zhang},\ and\ \citenamefont {Guo}}]{Zhang2010-1112}%
  \BibitemOpen
  \bibfield  {author} {\bibinfo {author} {\bibfnamefont {J.}~\bibnamefont
  {Zhang}}, \bibinfo {author} {\bibfnamefont {H.}~\bibnamefont {Yang}},
  \bibinfo {author} {\bibfnamefont {G.}~\bibnamefont {Shen}}, \bibinfo {author}
  {\bibfnamefont {P.}~\bibnamefont {Cheng}}, \bibinfo {author} {\bibfnamefont
  {J.}~\bibnamefont {Zhang}}, \ and\ \bibinfo {author} {\bibfnamefont
  {S.}~\bibnamefont {Guo}},\ }\href@noop {} {\bibfield  {journal} {\bibinfo
  {journal} {Chem. Commun.}\ }\textbf {\bibinfo {volume} {46}},\ \bibinfo
  {pages} {1112} (\bibinfo {year} {2010})}\BibitemShut {NoStop}%
\bibitem [{\citenamefont {Dietercih}\ \emph {et~al.}(1977)\citenamefont
  {Dietercih}, \citenamefont {Peschel},\ and\ \citenamefont
  {Schneider}}]{Dieterich1977-177}%
  \BibitemOpen
  \bibfield  {author} {\bibinfo {author} {\bibfnamefont {W.}~\bibnamefont
  {Dietercih}}, \bibinfo {author} {\bibfnamefont {I.}~\bibnamefont {Peschel}},
  \ and\ \bibinfo {author} {\bibfnamefont {W.~R.}\ \bibnamefont {Schneider}},\
  }\href@noop {} {\bibfield  {journal} {\bibinfo  {journal} {Z. Physik B}\
  }\textbf {\bibinfo {volume} {27}},\ \bibinfo {pages} {177} (\bibinfo {year}
  {1977})}\BibitemShut {NoStop}%
\bibitem [{\citenamefont {Kubo}\ \emph {et~al.}(1985)\citenamefont {Kubo},
  \citenamefont {Toda},\ and\ \citenamefont {Hashitsume}}]{Kubo1985}%
  \BibitemOpen
  \bibfield  {author} {\bibinfo {author} {\bibfnamefont {R.}~\bibnamefont
  {Kubo}}, \bibinfo {author} {\bibfnamefont {M.}~\bibnamefont {Toda}}, \ and\
  \bibinfo {author} {\bibfnamefont {N.}~\bibnamefont {Hashitsume}},\
  }\href@noop {} {\emph {\bibinfo {title} {Statistical Physics II:
  Nonequilibrium Statistical Mechanics}}},\ Solid-State Sciences\ (\bibinfo
  {publisher} {Springer-Verlag},\ \bibinfo {address} {Berlin},\ \bibinfo {year}
  {1985})\BibitemShut {NoStop}%
\bibitem [{\citenamefont {Mahan}(2000)}]{Mahan2000}%
  \BibitemOpen
  \bibfield  {author} {\bibinfo {author} {\bibfnamefont {G.~D.}\ \bibnamefont
  {Mahan}},\ }\href@noop {} {\emph {\bibinfo {title} {Many-Particle
  Physics}}},\ \bibinfo {edition} {3rd}\ ed.\ (\bibinfo  {publisher} {Kluwer
  Academic/Plenum Publishers},\ \bibinfo {address} {New York},\ \bibinfo {year}
  {2000})\BibitemShut {NoStop}%
\bibitem [{\citenamefont {Braun}\ and\ \citenamefont
  {Sholl}(1998)}]{Braun1998-14870}%
  \BibitemOpen
  \bibfield  {author} {\bibinfo {author} {\bibfnamefont {O.~M.}\ \bibnamefont
  {Braun}}\ and\ \bibinfo {author} {\bibfnamefont {C.~A.}\ \bibnamefont
  {Sholl}},\ }\href@noop {} {\bibfield  {journal} {\bibinfo  {journal} {Phys.
  Rev. B}\ }\textbf {\bibinfo {volume} {58}},\ \bibinfo {pages} {14870}
  (\bibinfo {year} {1998})}\BibitemShut {NoStop}%
\end{thebibliography}

%

\end{document}